\documentclass[12pt,english]{article}

\usepackage[T1]{fontenc}
\usepackage[utf8]{inputenc}
\usepackage{geometry}
\usepackage{amsmath}
\usepackage{amssymb}
\usepackage{graphicx}
\usepackage{setspace}

\makeatletter

\providecommand{\tabularnewline}{\\}

\newcommand{\lyxaddress}[1]{
	\par {\raggedright #1
	\vspace{1.4em}
	\noindent\par}
}

\usepackage[winfonts,UTF8]{ctex}

\usepackage{slashed}
\usepackage{hyperref}
\usepackage{graphicx}
\usepackage[numbers,sort&compress]{natbib}
\date{}

\hypersetup{colorlinks=true,citecolor=blue,linkcolor=red}

\makeatother

\usepackage{babel}
\begin{document}
\title{Evolution of Anti-de Sitter black holes in Einstein-Maxwell-dilaton theory}
\author{Cheng-Yong Zhang$^{1}$, Peng Liu$^{1}$, Yunqi Liu$^{2}$, Chao
Niu$^{1}$, Bin Wang$^{2,3}$ \thanks{zhangcy@email.jnu.edu.cn, phylp@email.jnu.edu.cn, yunqiliu@yzu.edu.cn, niuchaophy@gmail.com,
wang\_b@sjtu.edu.cn}}
\maketitle

\lyxaddress{\begin{center}
\textit{1. Department of Physics and Siyuan Laboratory, Jinan University,
Guangzhou 510632, China}\\
\textit{2. Center for Gravitation and Cosmology, College of Physical
Science and Technology, Yangzhou University, Yangzhou 225009, China}\\
\textit{3. School of Aeronautics and Astronautics, Shanghai Jiao Tong
University, Shanghai 200240, China}
\par\end{center}}
\begin{abstract}
  
  We study the nonlinear evolution of the spherical symmetric black holes under a small neutral scalar field perturbation in Einstein-Maxwell-dilaton theory with coupling function $f(\phi)=e^{-b\phi}$ in asymptotic anti-de Sitter spacetime. The non-minimal coupling between scalar and Maxwell fields allows the transmission of the energy from the Maxwell field to the scalar field, but also behaves as a repulsive force for the scalar. The scalar field oscillates with damping amplitude and converges to a final value by a power law. The irreducible mass of the black hole increases abruptly at initial times and then saturates to the final value exponentially. The saturating rate is twice the decaying rate of the dominant mode of the scalar. The effects of the black hole charge, the cosmological constant and the coupling parameter on the evolution are studied in detail. When the initial configuration is a naked singularity spacetime with a large charge to mass ratio, a horizon will form soon and hide the singularity.

\end{abstract}

\section{Introduction}

The famous no-hair theorem in Einstein-Maxwell theory and its generalisation in scalar-tensor theory shows that a black hole can be completely characterized by three degrees of freedom: its mass, charge and angular momentum \cite{Israel1967,Carter1971,Ruffini1971,HHBekenstein1995,Sotiriou:2011dz,Hui:2012qt}.
However, there are exceptions to the rule as well.
In gravitational theories beyond General Relativity (GR) the dilatonic and colored black holes in the Einstein-dilaton-Gauss-Bonnet theory \cite{Torii1997,Kanti1997} and the rotating \cite{EdGBP1,EdGBP2,EdGBP3} or higher dimensional \cite{EdGBP4,EdGBP5,EdGBP6,EdGBP7,EdGBP8,EdGBP9} or shift-symmetric Galileon \cite{EdGBP10,EdGBP11,EdGBP12} hairy black hole solutions circumvent the no-hair theorem.
Or even GR with certain matter sources could evade the no-hair theorem, such as black hole solutions with a Yang-Mills field \cite{HHVolkov1989,HHBizon1990,HHGreene1993,HHMaeda1994}, Skyrme field \cite{HHLuckock1986,HHDroz1991} and a conformally-coupled scalar field \cite{HHBekenstein1975}.
Recently, as a dynamic mechanism leading to hairy black hole solutions under the frame of GR, spontaneous scalarization has attracted much attention.
It was initially proposed in the study of the neutron star and also happens when black holes are surrounded by enough matter in scalar-tensor theory \cite{Damour1993,Cardoso1305,Cardoso1308,Zhang:2014kna}.
Recent studies show that spontaneous scalarization could exists typically in the models containing non-minimally coupling of a real scalar field to the source terms which could either be the geometric invariant sources such as the Ricci scalar, Gauss-Bonnet, Chern-Simons invariant \cite{Herdeiro:2019yjy,Doneva1711,Silva1711,Antoniou1711,Cunha1904,Dima:2020yac,Herdeiro2009,Berti2009,Lin:2020asf,Guo:2020sdu,Collodel:2019kkx,Doneva:2018rou,Brihaye:2018bgc} or the matter invariant sources such as Maxwell invariant in Einstein-Maxwell-scalar (EMS) theory \cite{Herdeiro:2018wub,Guo:2021zed,Blazquez-Salcedo:2020nhs,Herdeiro:2019iwl,Astefanesei:2020qxk,Astefanesei:2019pfq,Fernandes:2019kmh} and Einstein-Maxwell-vector model \cite{Oliveira:2020dru}.
The successful observation of gravitational waves \cite{Abbott1,Abbott:2016blz,Barack:2018yly} and the black hole shadow \cite{Akiyama:2019cqa,Akiyama:2019bq,Akiyama:2019eap} have pushed the research and theoretical detection of black holes into a new era providing a new window to test the characteristics of black holes, thus it is necessary to carefully study the physics on hairy black holes such as metrics, the dynamic process of formation and evolution, and use the observation to test no-hair theorem or constraint the families of black holes\cite{Khodadi:2020jij,Rahmani:2020vvv}.

As a natural and simple generalisation of the Einstein-Maxwell theory, the Einstein-Maxwell-dilaton (EMD) theory occurs in the context of Kaluza-Klein theories in which the scalar field describes how the extra dimensions dilate along the four-dimensional spacetime \cite{Kaluza:1921tu} It also originates from the low energy limit of string theory and is ubiquitous in supergravity \cite{Cremmer:1978ds}. The dilaton couples to the Maxwell term non-minimally and prevents the Reissner-Nordstr\"om (RN) solution in EMD theory. Instead, only dilatonic black hole solutions exist that present some RN-unlike features such as that the black hole charge to mass ratio can exceed unity  \cite{Gibbons:1987ps,Garfinkle:1990qj,Zhang:2015jda}. The solutions here have mass, charge, rotation, and scalar hair, together with scalar, vector, and tensor radiative channels, and therefore offer an interesting theoretical and computational playground to explore possible deviations from the general relativity prediction.

The mathematical structure of the EMD model allows for defining a well posed initial value problem rather than those in the  extended scalar-tensor-Gauss-Bonnet models \cite{Ripley:2019irj,Ripley:2019aqj,Ripley:2020vpk,Doneva:2021dqn,Silva:2020omi,Witek:2018dmd}.
In asymptotic flat spacetime for EMD theory, the black hole endowed with a potential emerging from low energy heterotic string theory was found nonlinearly stable under perturbations \cite{Astefanesei:2019qsg}. The dynamical evolution of individual black holes, as well as the merger of binary black hole systems, were analyzed in \cite{Hirschmann:2017psw}. They found that the black hole systems are difficult to be distinguished from their analogs within general relativity when the charge is relatively small. The dynamics in EMS models with more generic non-minimally coupling functions are also of interest, especially from the viewpoint of spontaneous scalarisation. Here the RN black hole is a solution but unstable against scalar perturbations for a sufficiently large charge to mass ratio. The scalar hairy black hole solution is energetically and dynamically favored \cite{Herdeiro:2018wub,Fernandes:2019kmh,Fernandes:2019rez}.

In this paper, we will study the evolution of an initial RN-anti-de Sitter (AdS) black hole under a small neutral scalar field perturbation in EMD theory. It is known that the cosmological constant can significantly affect the scalarisation of the black hole \cite{Zhang:2014kna}. The regular hairy black hole solutions exist in all asymptotic flat, de Sitter (dS) and anti-dS (AdS) spacetime in EMS model \cite{Guo:2021zed,Brihaye:2019gla,Zhang:2021etr}. While in the extended scalar-tensor-Gauss-Bonnet model, they exist only in asymptotic flat and AdS spacetime \cite{Bakopoulos:2018nui}. The positive cosmological constant can quench the tachyonic instability. While the dynamically and thermodynamically stable scalarised black holes can exist in EMD theory \cite{Astefanesei:2019qsg}. Therefore, it is necessary to study the full non-linear dynamical evolution of the black holes in asymptotic AdS spacetime in EMD theory and compare the results with those in asymptotic flat spacetime. Note that the dynamics in AdS spacetime is qualitatively different from those in asymptotic flat spacetime since the scalar modes can propagate to the spacial boundary in finite coordinate time and be bounced back. The studies in asymptotical AdS spacetime can not be naively generalized to the case of asymptotically flat spacetime since their boundary behaviors are distinct.

This paper is organized as follows. In section \ref{sec:emd}, we introduce the
equations of motions and boundary behaviors of the variables in EMD
theory. In section \ref{sec:num}, we demonstrate the numerical results, where the effects of the coupling function parameter $b$ (\ref{subsec:b}), the charge (\ref{subsec:q}) and the cosmological constant (\ref{subsec:l}) on the dynamical scalarisation are studied in detail. Section \ref{sec:sum} gives the summery and discussion.

\section{Einstein-Maxwell-dilaton  theory}\label{sec:emd}

The action of Einstein-Maxwell-dilaton theory in AdS spacetime in this work is
\begin{equation}
S=\frac{1}{16\pi}\int d^{4}x\sqrt{-g}\left[R-2\Lambda-2\nabla_{\mu}\phi\nabla^{\mu}\phi-e^{-b\phi}F_{\mu\nu}F^{\mu\nu}\right].
\end{equation}
The cosmological constant $\Lambda$ is negative for asymptotic AdS spacetime. $R$ is the Ricci scalar of the metric $g_{\mu\nu}$. The Maxwell field strength is $F_{\mu\nu}=\partial_{\mu}A_{\nu}-\partial_{\nu}A_{\mu}$ in which $A_{\mu}$ is the gauge field. The coupling function between the real dilaton $\phi$ and gauge field is $f(\phi)=e^{-b\phi}$ in which the parameter $b$ is a constant. Note that the action is invariant under the $\mathbb{Z}_{2}$ symmetry $(b,\phi)\to-(b,\phi)$. Hereafter, we keep $-b>0$ in this paper without loss of generality. The dilatonic coupling function appears in Kaluza-Klein models, supergravity or low-energy string models. Some of the exact solutions in asymptotic flat spacetime are obtained in \cite{Gibbons:1987ps,Garfinkle:1990qj}.

The equations of motion are
\begin{align}
R_{\mu\nu}-\frac{1}{2}Rg_{\mu\nu}+\Lambda g_{\mu\nu}= & 2\left[\partial_{\mu}\phi\partial_{\nu}\phi-\frac{1}{2}g_{\mu\nu}\nabla_{\rho}\phi\nabla^{\rho}\phi+e^{-b\phi}\left(F_{\mu\rho}F_{\nu}^{\ \rho}-\frac{1}{4}g_{\mu\nu}F_{\rho\sigma}F^{\rho\sigma}\right)\right],\\
\nabla_{\mu}\nabla^{\mu}\phi= & -\frac{b}{4} e^{-b\phi}F_{\mu\nu}F^{\mu\nu},\label{eq:phi}\\
\nabla_{\mu}\left(f(\phi)F^{\mu\nu}\right)= & 0.
\end{align}
It is obvious that $\phi=0$ will not be a solution of (\ref{eq:phi}) unless $A_\mu=0$ and thus the RN-AdS black hole is not a solution of EMD theory. To study the dynamic evolution of the black hole, we take the ingoing Eddington-Finkelstein coordinate ansatz
\begin{equation}
ds^{2}=-\alpha(t,r)dt^{2}+2dtdr+\zeta(t,r)^{2}(d\theta^{2}+\sin^{2}\theta d\phi^{2}).
\end{equation}
The coordinate is regular on the black hole apparent horizon which
is defined by
\begin{equation}
0=g^{ab}\partial_{a}\zeta\partial_{b}\zeta.
\end{equation}
Here indexes ${a,b}\in\{t,r\}.$
Once we get the apparent horizon $r_{a}$, we can get the irreducible mass of the black hole $M_{0}=\sqrt{\frac{A}{4\pi}}=\zeta(t,r_{a})$ in which $A=4\pi\zeta^{2}(t,r_{a})$ is the area of the black hole. As the black hole area does not decrease, we will see that the irreducible mass of the black hole will not decrease in the evolution. On the other hand, we will study the rescaled Misner-Sharp mass defined as $M_{MS}=m/4\pi$ in which the generalized Misner-Sharp quasi-local mass is defined by \cite{Maeda:2012fr}
\begin{equation}
m=2\pi\zeta\left(-\frac{\Lambda}{3}\zeta^{2}+1-g^{ab}\partial_a\zeta\partial_{b}\zeta\right).
\end{equation}
The rescaled Misner-Sharp mass tends to the ADM mass of the spacetime as $r\to\infty$.

We also require the gauge potential $A_{\mu}dx^{\mu}=A(t,r)dt$ and
the dilaton $\phi=\phi(t,r)$. With these choices, the Maxwell field
can be worked out as
\begin{equation}\label{eq:electric}
\partial_{r}A=\frac{Q}{\zeta^{2}f(\phi)},
\end{equation}
in which $Q$ is a constant interpreted as the electric charge of the black hole.
Eq.(\ref{eq:electric}) indicates the strength of Maxwell field.
The coupling function $f(\phi)$ acts as an effective dielectric  that varying the strength.
To implement the numerics, we introduce auxiliary variables
\begin{equation}
S=\partial_{t}\zeta+\frac{1}{2}\alpha\partial_{r}\zeta,\ \ \ \ P=\partial_{t}\phi+\frac{1}{2}\alpha\partial_{r}\phi.\label{eq:Pt}
\end{equation}
Then the Einstein equations become
\begin{align}
\partial_{t}S=& \frac{1}{2}S\partial_{r}\alpha+\frac{\alpha}{2}\left(\frac{2S\partial_{r}\zeta-1}{2\zeta}+\frac{1}{2}\zeta\Lambda+\frac{Q^{2}}{2\zeta^{3}f(\phi)}\right)-\zeta P^{2},\label{eq:St}\\
\partial_{r}^{2}\alpha= & -4P\partial_{r}\phi+\frac{4S\partial_{r}\zeta-2}{\zeta^{2}}+\frac{4Q^{2}}{\zeta^{4}f(\phi)},\label{eq:alphar}\\
\partial_{r}S= & \frac{1-2S\partial_{r}\zeta}{2\zeta}-\frac{\zeta\Lambda}{2}-\frac{Q^{2}}{2\zeta^{3}f(\phi)},\label{eq:Sr}\\
\partial_{r}^{2}\zeta= & -\zeta(\partial_{r}\phi){}^{2}.\label{eq:zetar}
\end{align}
The scalar equation becomes
\begin{equation}
P'=-\frac{P\zeta'+S\phi'}{\zeta}-\frac{Q^{2}}{4\zeta^{4}f(\phi)^{2}}\frac{df(\phi)}{d\phi}.\label{eq:Pr}
\end{equation}
Given an initial $\phi$, we can integrate (\ref{eq:alphar},\ref{eq:Sr},\ref{eq:zetar},\ref{eq:Pr}) to get $\alpha,S,\zeta,P$ at the initial time. Then from (\ref{eq:Pt}) we get the $\phi$ at the next time step. (\ref{eq:St}) is redundant and can be used to check the accuracy of the numerical code. To solve these equations numerically, we need to specify boundary conditions. Expanding the variables in the asymptotic infinity, we get the asymptotic
solutions
\begin{align}
\phi= & \frac{\phi_{3}(t)}{r^{3}}+\frac{3}{8\Lambda r^{4}}\left(-bQ^2-8\phi'_{3}(t)\right)+O(r^{-5}),\\
\alpha= & -\frac{\Lambda}{3}r^{2}+1-\frac{2M}{r}+\frac{Q^{2}}{r^{2}}+\frac{\Lambda}{5r^{4}}\phi_{3}^{2}(t)+O(r^{-5}),\\
\zeta= & r-\frac{3\phi_{3}^{2}(t)}{10r^{5}}+\frac{3\phi_{3}(t)}{14\Lambda r^{6}}\left(-bQ^2-8\phi'_{3}(t)\right)+O(r^{-7}),\\
S= & -\frac{\Lambda}{6}r^{2}+\frac{1}{2}-\frac{M}{r}+\frac{Q^{2}}{2r^{2}}-\frac{3\Lambda}{20r^{4}}\phi_{3}^{2}(t)+O(r^{-5}),\\
P= & \frac{\Lambda\phi_{3}(t)}{2r^{2}}+\frac{1}{r^{3}}\left(\frac{-bQ^{2}}{4}-\phi'_{3}(t)\right)+\frac{3}{2\Lambda r^{4}}\phi''_{3}(t)+O(r^{-5}),
\end{align}
in which  $\phi'_{3}(t)=\frac{d\phi_{3}(t)}{dt}.$
The free parameters of the asymptotic solution are the ADM mass $M$ and charge $Q$ of
the black hole, and the cosmological constant $\Lambda$. Function
$\phi_{3}(t)$ is unknown and should be determined by evolution.
At static case, $\phi_{3}$ can be viewed as the parameter indicating the existence of the scalar hair. Note that we have set $\zeta-r=0$
as $r\to\infty$ by fixing the residual radial reparameterization
freedom \cite{Chesler:2013lia}. Some variables such as $\zeta,\alpha$
and $S$ are divergent at infinity. We introduce the following new
variables to do the numerical calculation:
\begin{equation}
\zeta\equiv r\sigma,\alpha\equiv r^{2}a,S\equiv r^{2}s,P\equiv\frac{1}{r}p.
\end{equation}

In asymptotic AdS spacetime, the scalar perturbation can reach the infinity in finite coordinate time and be bounced back to the bulk. We must include the infinity in the computational domain. We thus compactify the radial direction by a coordinate transformation: \begin{equation}
z=\frac{r}{r+M}.
\end{equation}
The computation domain in $z$ coordinate is then $(z_{i},1)$ where $z_{i}$ corresponds to some radius $r_{i}$ which is close to the black hole horizon from inside and $z=1$ corresponds to radial infinity. Now the boundary conditions at $z=1$ are
\begin{align}
\sigma=1,\sigma'=0, & s=-\frac{\Lambda}{6},s'=0,s''=6(M-1),\\
p=0, & a=-\frac{\Lambda}{3},a'=0,a''=12(M-1).
\end{align}
The $z$ direction is uniformly discretized. Equations (\ref{eq:alphar},\ref{eq:zetar},\ref{eq:Pr}) are discretized with fourth-order finite difference while (\ref{eq:Sr}) is discretized with second-order finite difference. The time direction marches with fourth-order Runge-Kutta method. We employ Kreiss-Oliger dissipation to stabilize the numerical evolution.

\section{Numerical results}\label{sec:num}
{\footnotesize{}}
\begin{figure*}
\begin{centering}
{\footnotesize{}}%
\begin{tabular}{cc}
{\footnotesize{}\includegraphics[width=0.4\textwidth]{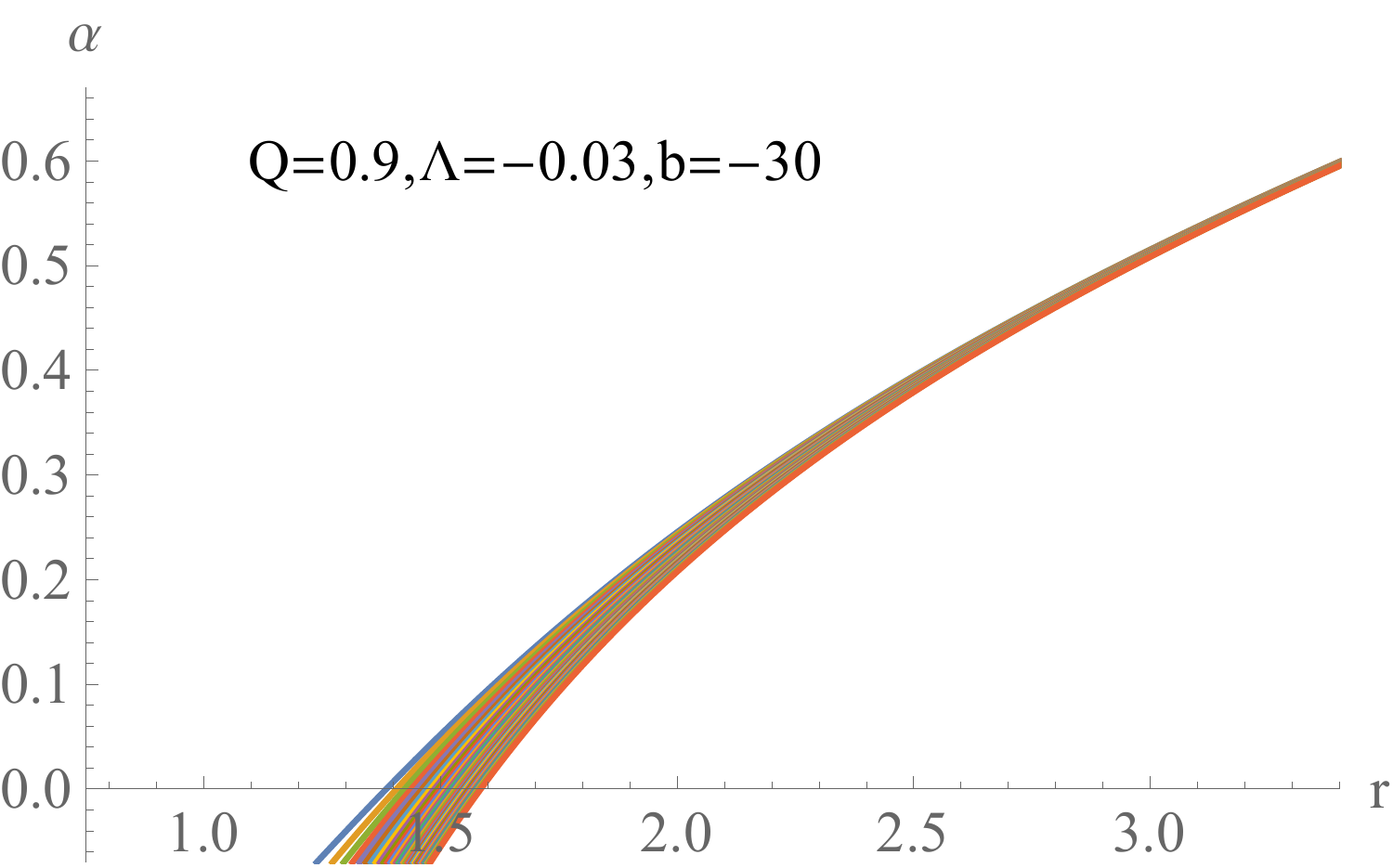}} & {\footnotesize{}\includegraphics[width=0.4\textwidth]{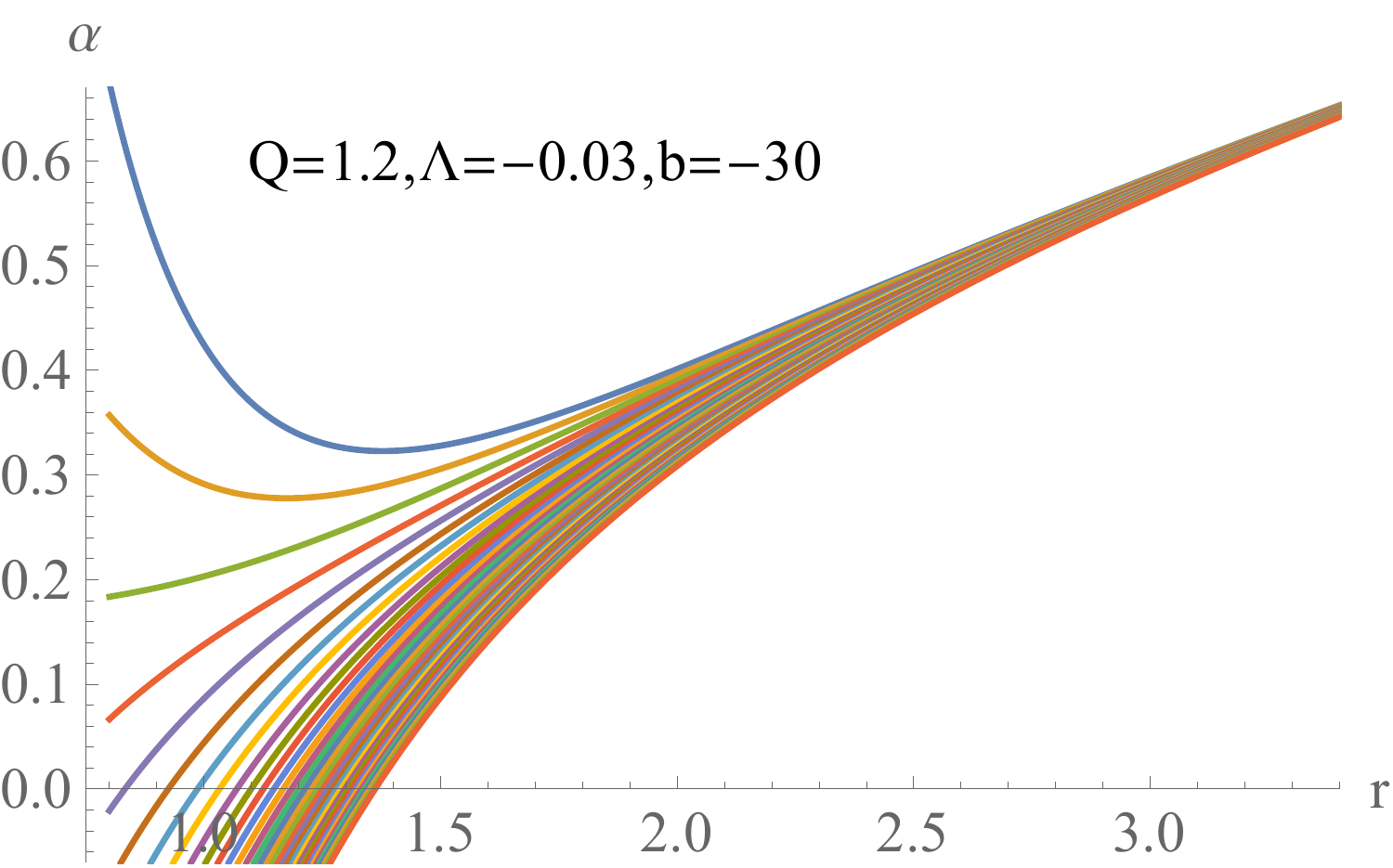}}\tabularnewline
{\footnotesize{}\includegraphics[width=0.4\textwidth]{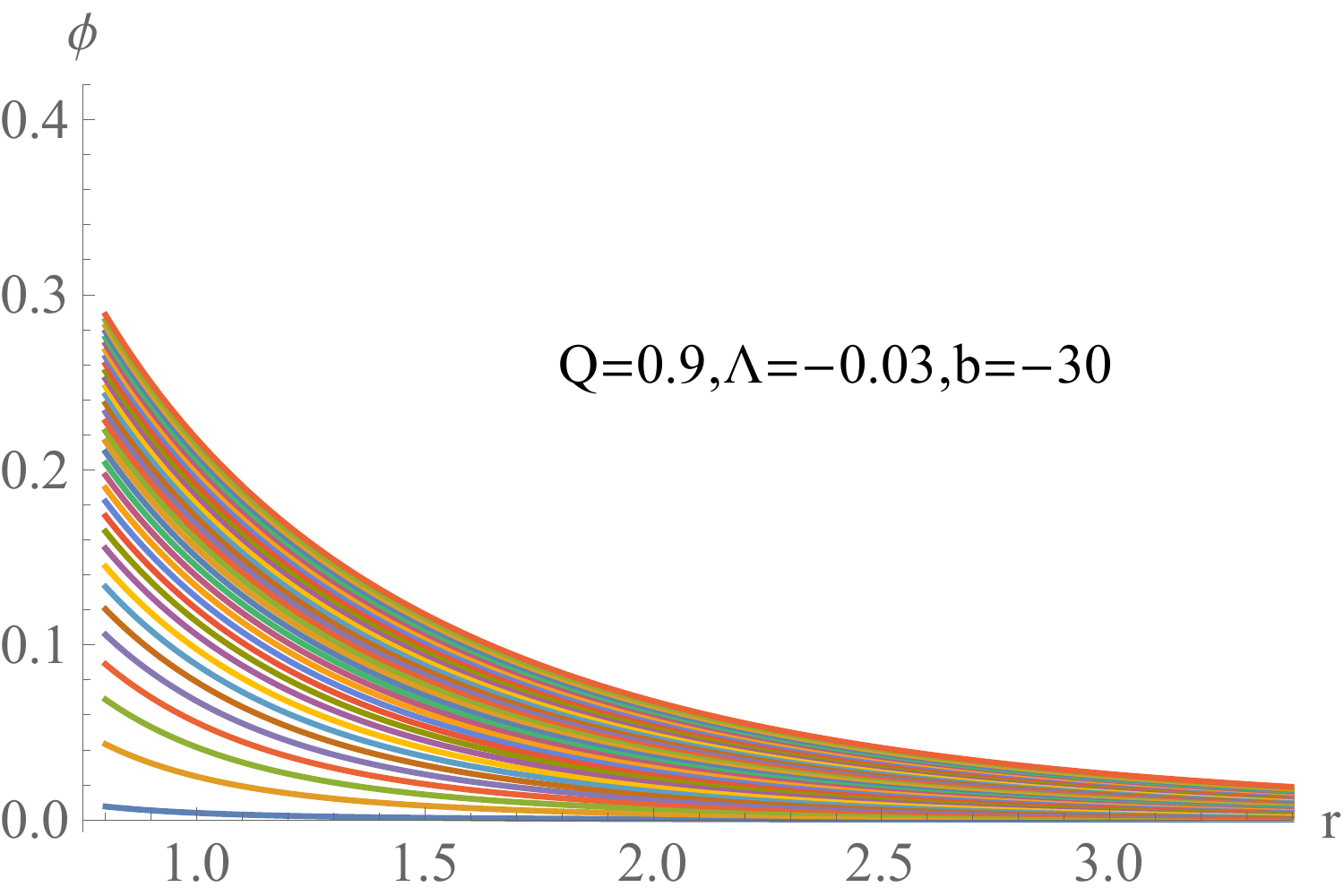}} & {\footnotesize{}\includegraphics[width=0.4\textwidth]{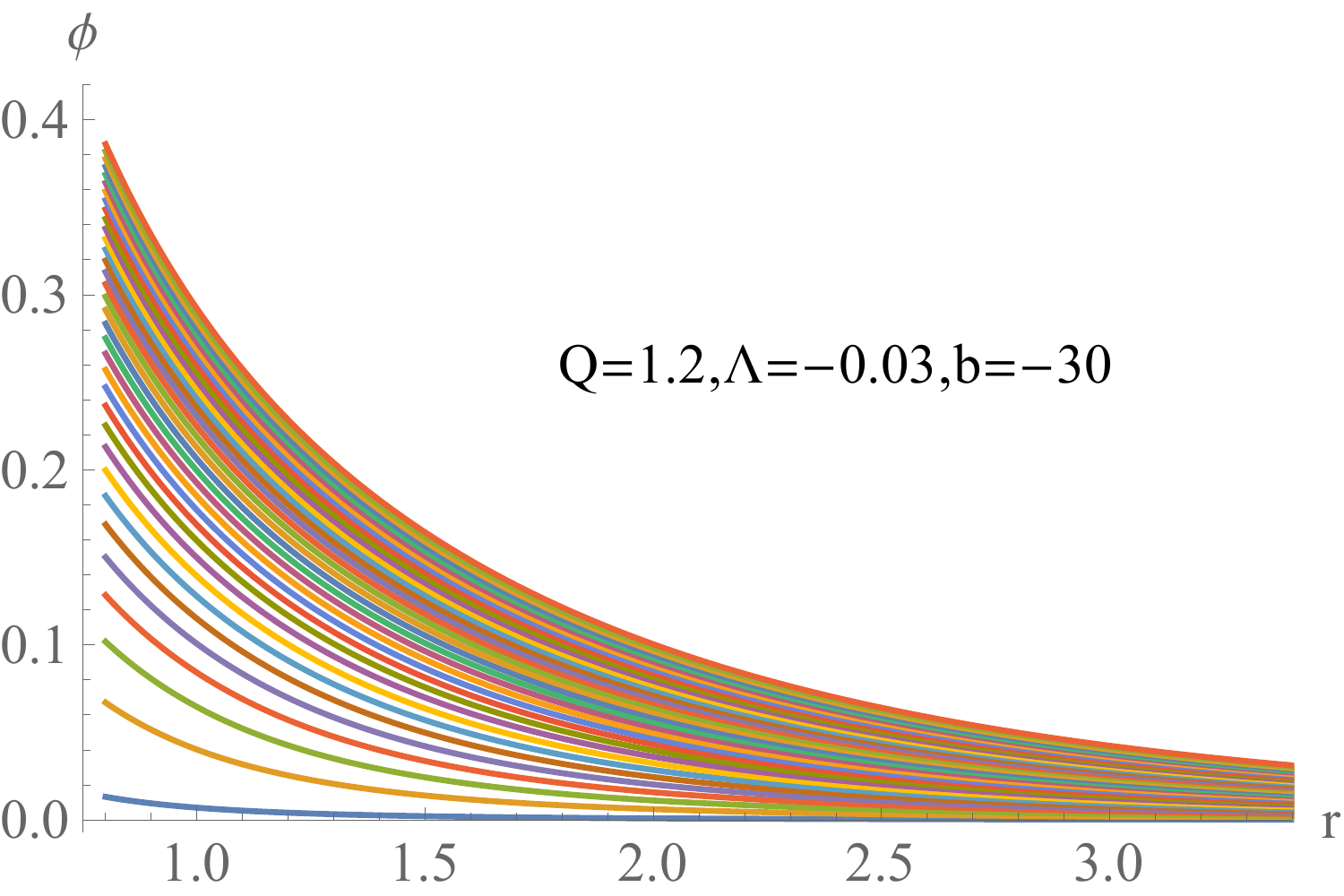}}\tabularnewline
\end{tabular}{\footnotesize\par}
\par\end{centering}
{\footnotesize{}\caption{\label{fig:BHsingular}The sketches of the very early time evolutions (from $t=0$ to $t=0.27$) of metric $\alpha$ and scalar $\phi$ starting from an initial black hole (left) or a naked singularity (right) spacetime. The blue line corresponds to the initial case. The time step between adjacent lines is $\Delta t\simeq0.0081$.}
}{\footnotesize\par}
\end{figure*}
{\footnotesize\par}
The free parameter of the system is the black hole charge $Q$, ADM
mass $M$ and the cosmological constant $\Lambda$. We take the initial
dilaton profile as
\begin{equation}
\phi_{0}=\kappa e^{-\frac{\left(r-4r_{h}\right)^{2}}{w^{2}}}\text{ or }\phi_{0}=\begin{cases}
\left(\frac{1}{r}-\frac{1}{r_{1}}\right)^{3}\left(\frac{1}{r}-\frac{1}{r_{2}}\right)^{3}\frac{\kappa_{1}+\kappa_{2}\sin\frac{10}{r}}{r^{2}}, & r_{1}<r<r_{2},\\
0 & r\le r_{1}\text{ or }r\ge r_{2}.
\end{cases}
\end{equation}
Here, $\kappa<10^{-9}$ and width $w=1.8r_{h}$ where $r_{h}$ is the
horizon of the corresponding RN-AdS black hole with  metric  $\alpha = 1-\frac{2M}{r}+\frac{Q^2}{r^2}-\frac{\Lambda r^2}{3}$. Initial parameters $\{r_{1},r_{2}\}=\{2r_{h},3r_{h}\}$
and $\kappa_{1},\kappa_{2}$ are of order $10^{-2}$ so that the initial
scalar field is of order $10^{-10}$ and negligible compared to the initial
black hole. {Hereafter, we fix $M=1$ in this paper to implement the dimensionless of the physical quantities. The critical charge $Q_c \simeq 1$ when $-\Lambda$ is small.

We show the sketches of the very early time evolutions of metric $\alpha$ and scalar $\phi$ in  Fig.\ref{fig:BHsingular}. When $Q<Q_c$, the initial configuration is a black hole spacetime. Part of the energy of the Maxwell field is transferred to the scalar field due to their non-minimal coupling. The scalar field grows rapidly and is captured by the black hole. So the scalar decreases monotonically in the radial direction and the radius of the apparent horizon increases with time. It is known that the charge of the hairy black hole solution in EMD theory could be greater than the corresponding $Q_c$ \cite{Garfinkle:1990qj}. We find that our numerical codes work well when $Q$ is greater than $Q_c$ but not too large. As shown in the right of Fig.\ref{fig:BHsingular}, the initial configuration is a spacetime with naked singularity. But a horizon forms soon and hides the singularity, resulting in a regular spacetime geometry outside the horizon. The scalar grows faster than the case when $Q<Q_c$.}

\subsection{Effects of coupling parameter $b$ on the black hole evolution}\label{subsec:b}

\begin{figure*}
\begin{centering}
{\footnotesize{}}%
\begin{tabular}{cc}
{\footnotesize{}\includegraphics[width=0.43\textwidth]{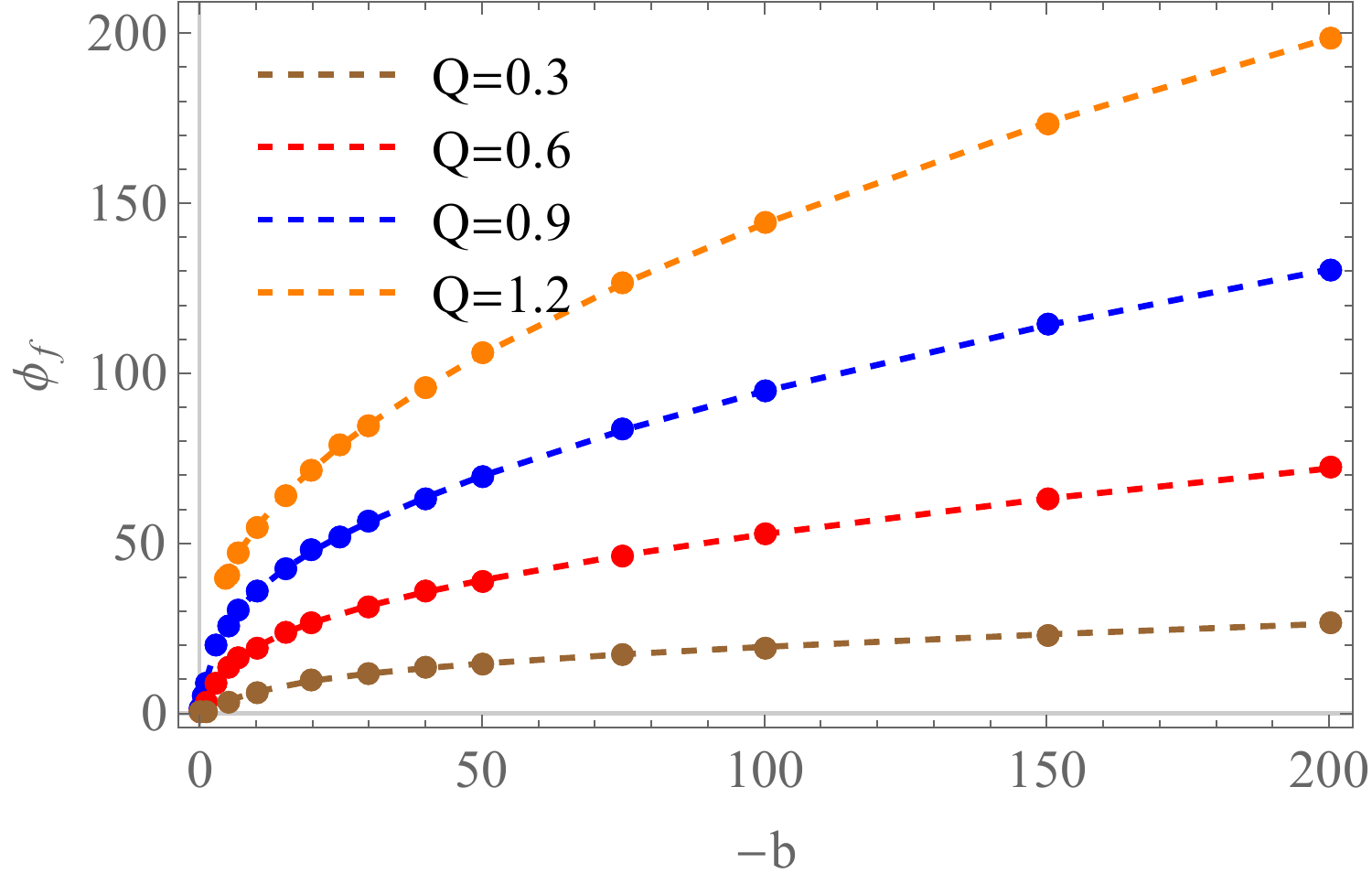}} & {\footnotesize{}\includegraphics[width=0.43\textwidth]{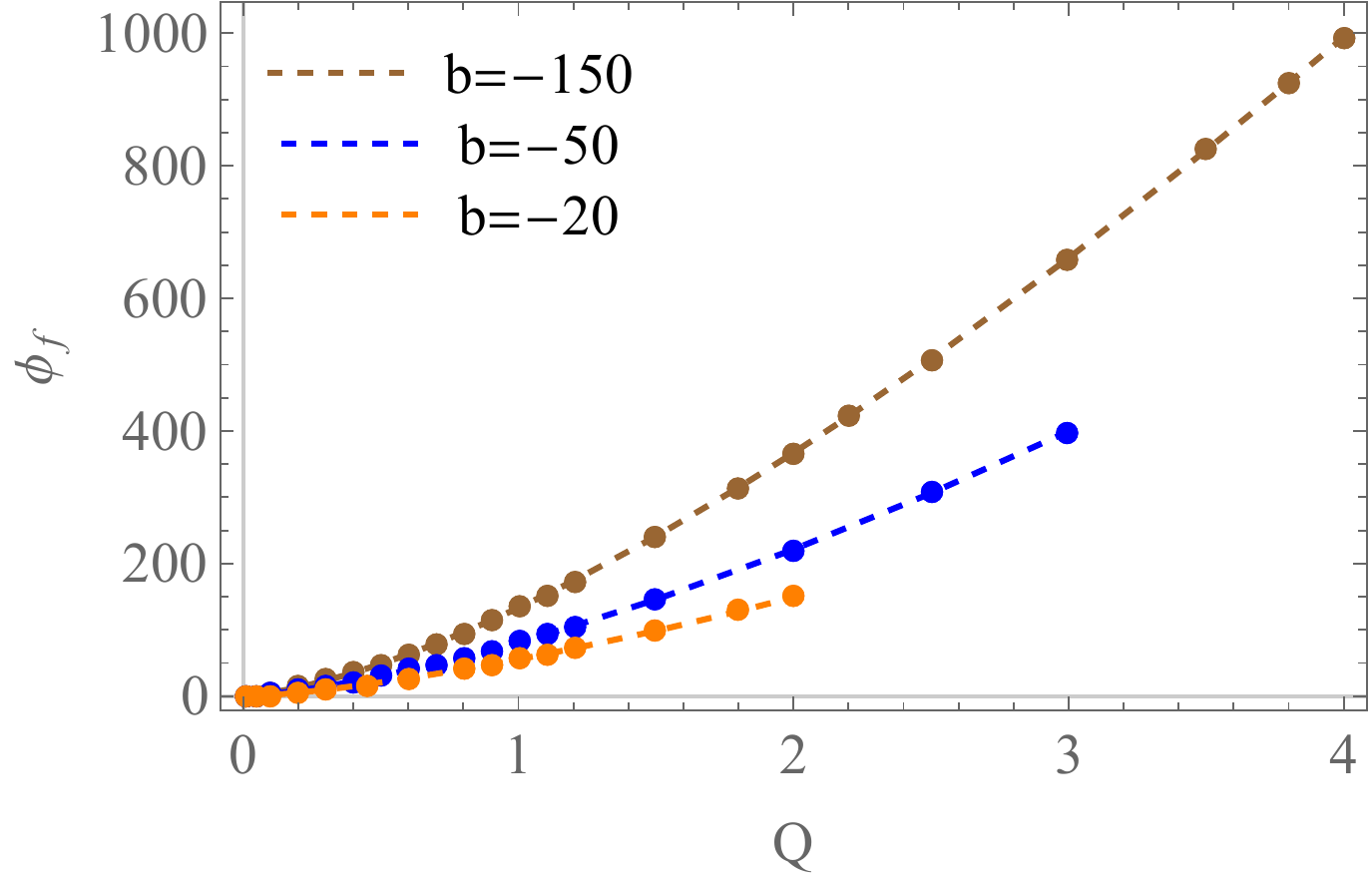}}\tabularnewline
\end{tabular}{\footnotesize\par}
\par\end{centering}
{\footnotesize{}\caption{\label{fig:Q963b2010phi3f}The final value $\phi_{f}$ of $\phi_{3}$
for various $b$ when $Q$ is fixed (left), and for various
$Q$ when $b$ is fixed (right). Note that in the left panel, regular hairy black hole solutions from the evolution of initial naked RN-AdS like singularity spacetimes exist for $-b>4.5$ when $Q=1.2$. In the right panel regular hairy black holes exist for $Q<2,3,4$ when $-b=20,50,150$, respectively.  Here $\Lambda=-0.03$.}
}{\footnotesize\par}
\end{figure*}

In this subsection, we fix $\Lambda=0.03$ and chose certain $Q$ to study the effects of coupling parameter $b$ on the black hole evolution.
Since $\phi_{3}$ can be viewed as an indicator of the scalar hair,
we show the final value $\phi_{f}$ of $\phi_{3}$ in the left panel of Fig.\ref{fig:Q963b2010phi3f}.
The $\phi_{f}$ increases with both $-b$ and $Q$.  Unlike the case
in EMS model where the static hairy black hole solution exists only
when $-b$ and $Q$ are large enough \cite{Zhang:2021etr}, there is always static hairy
black hole solution here when $Q<Q_c$. This is reasonable since only a hairy black hole solution exists in EMD theory. The initial  RN-AdS black hole solution is dressed with scalar hair soon after the evolution. As the non-minimal coupling between the scalar field and the Maxwell field becomes stronger, the black hole will be dressed heavier. When $Q>Q_c$, our numerical codes work well only if $-b$ is large. This implies that the weak non-minimal coupling between the scalar field and the Maxwell field can not destroy the original naked RN-AdS like singularity. Only when the coupling is strong enough, would the original naked singularity be destroyed, and the hairy black hole solution gradually develops.

{\footnotesize{}}
{\footnotesize\par}

\begin{figure*}
\begin{centering}
{\footnotesize{}}%
\begin{tabular}{cc}
{\footnotesize{}\includegraphics[width=0.41\textwidth]{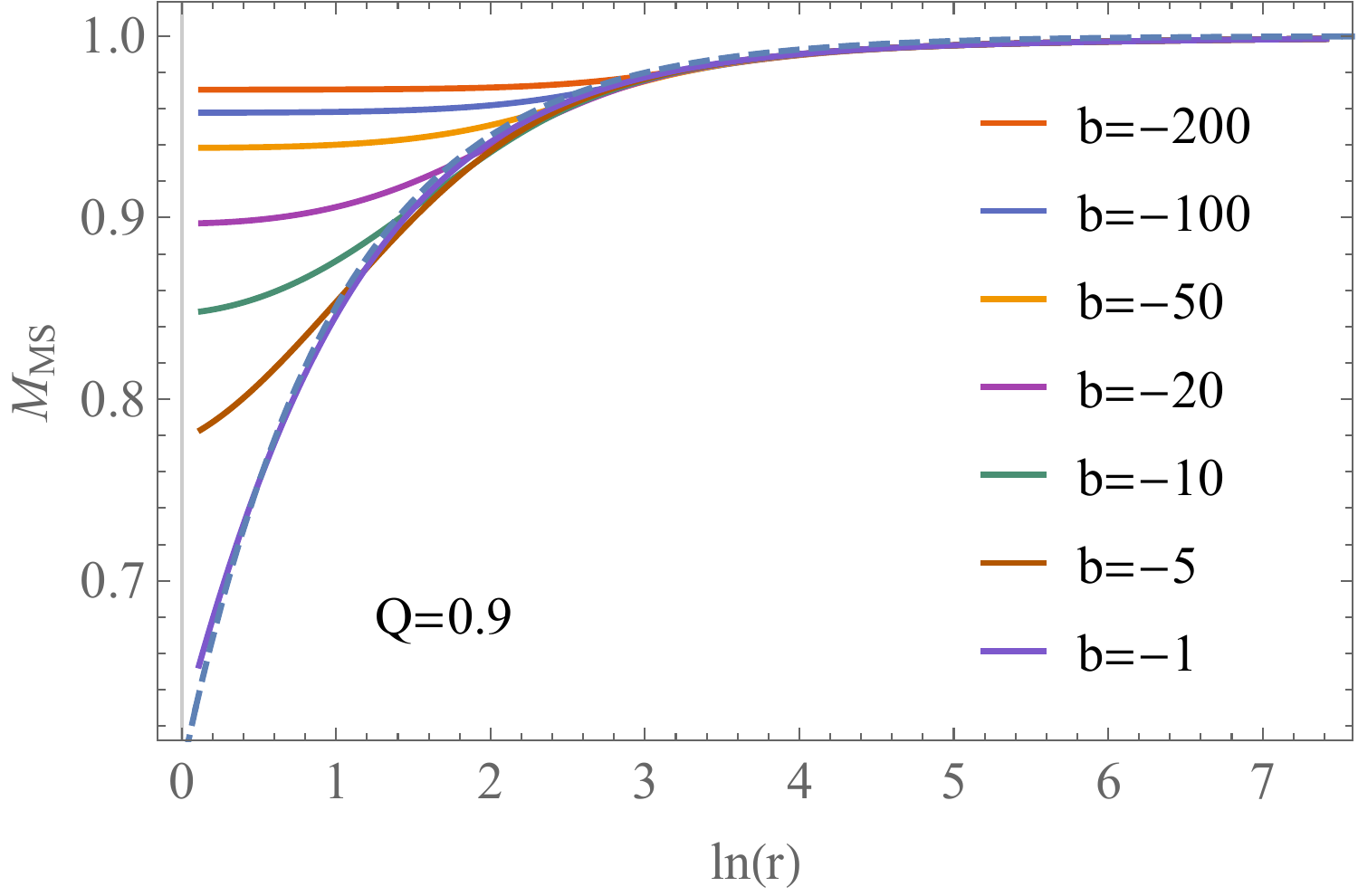}} & {\footnotesize{}\includegraphics[width=0.43\textwidth]{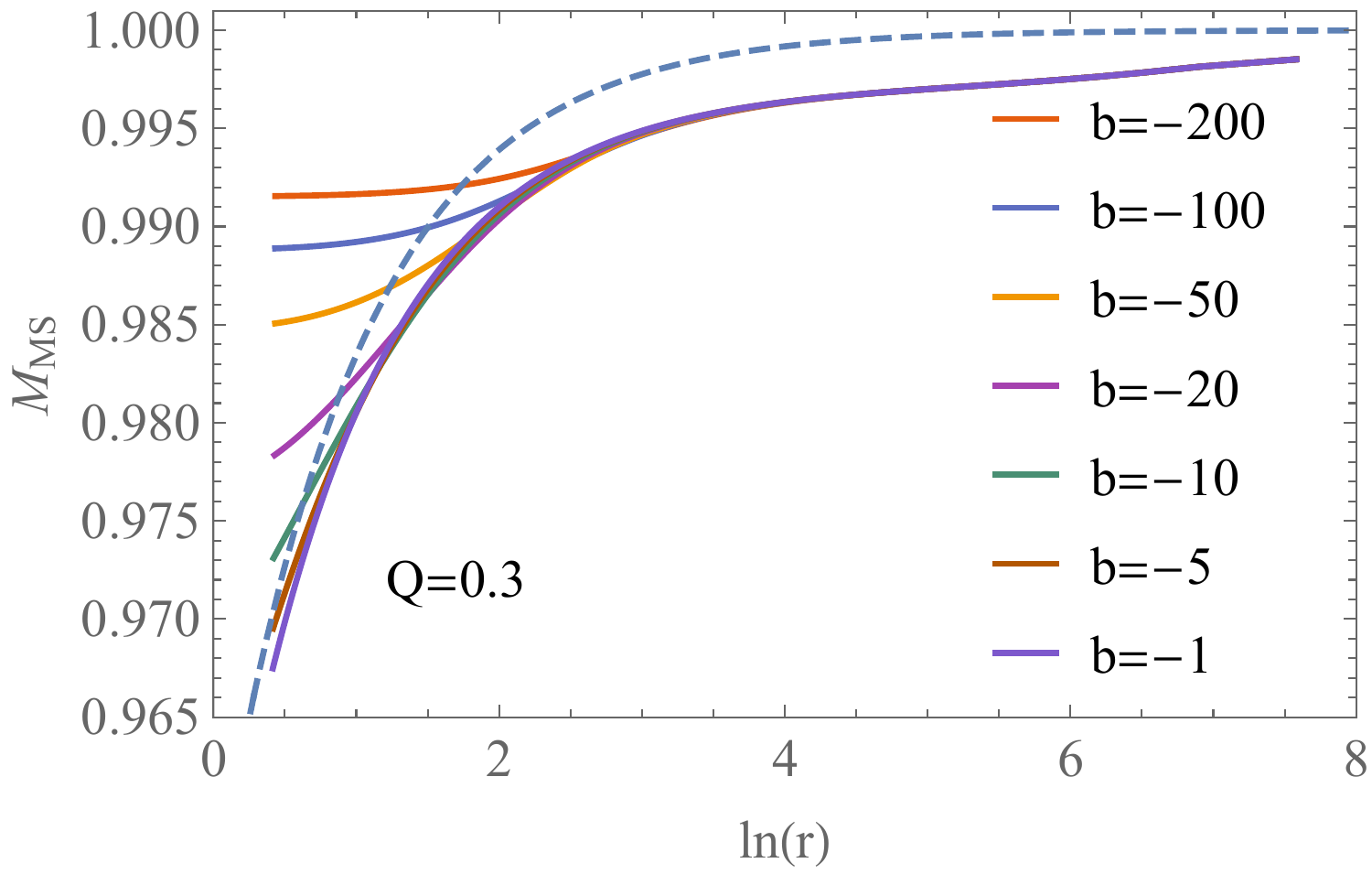}}\tabularnewline
{\footnotesize{}\includegraphics[width=0.41\textwidth]{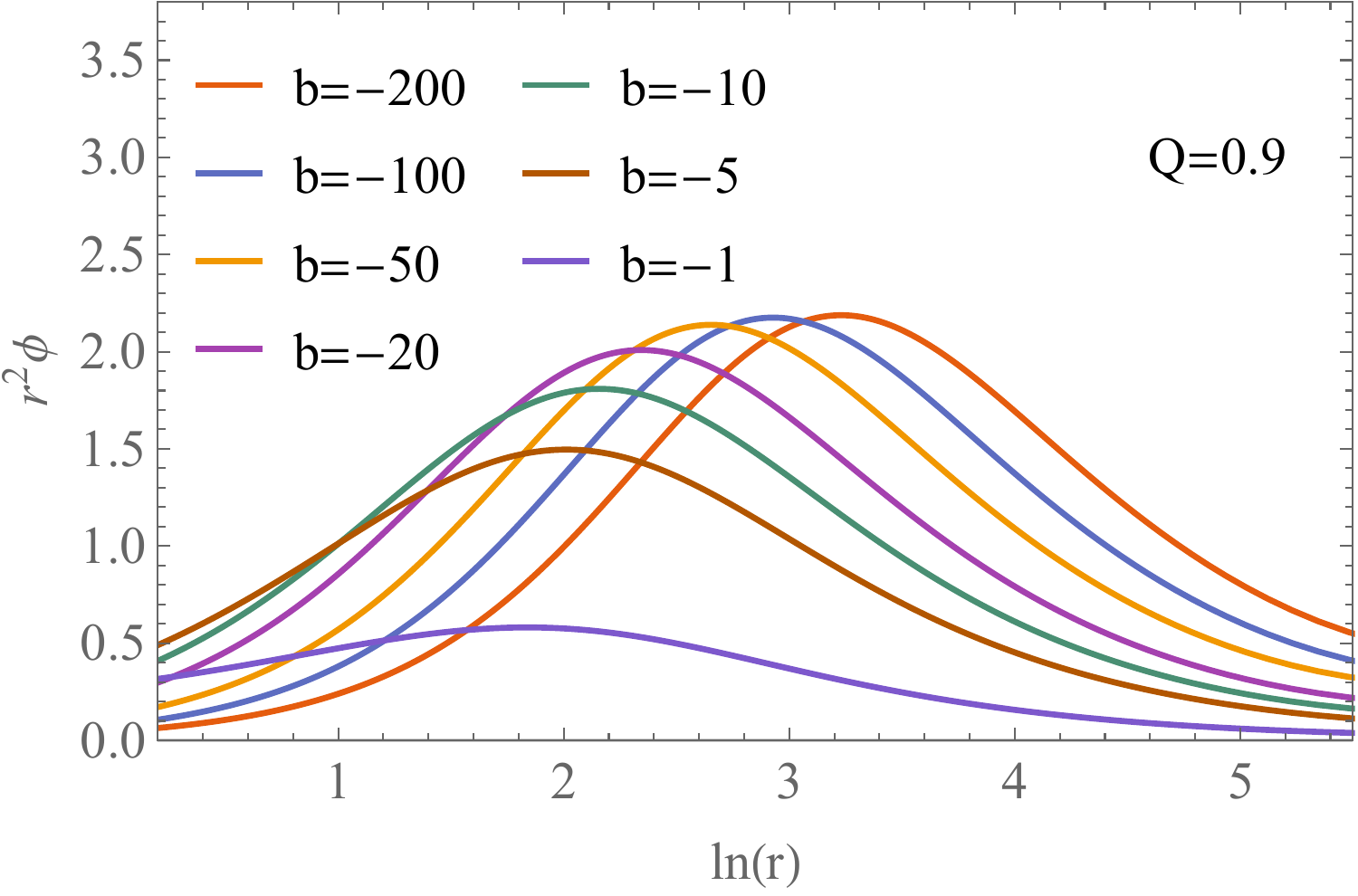}} & {\footnotesize{}\includegraphics[width=0.42\textwidth]{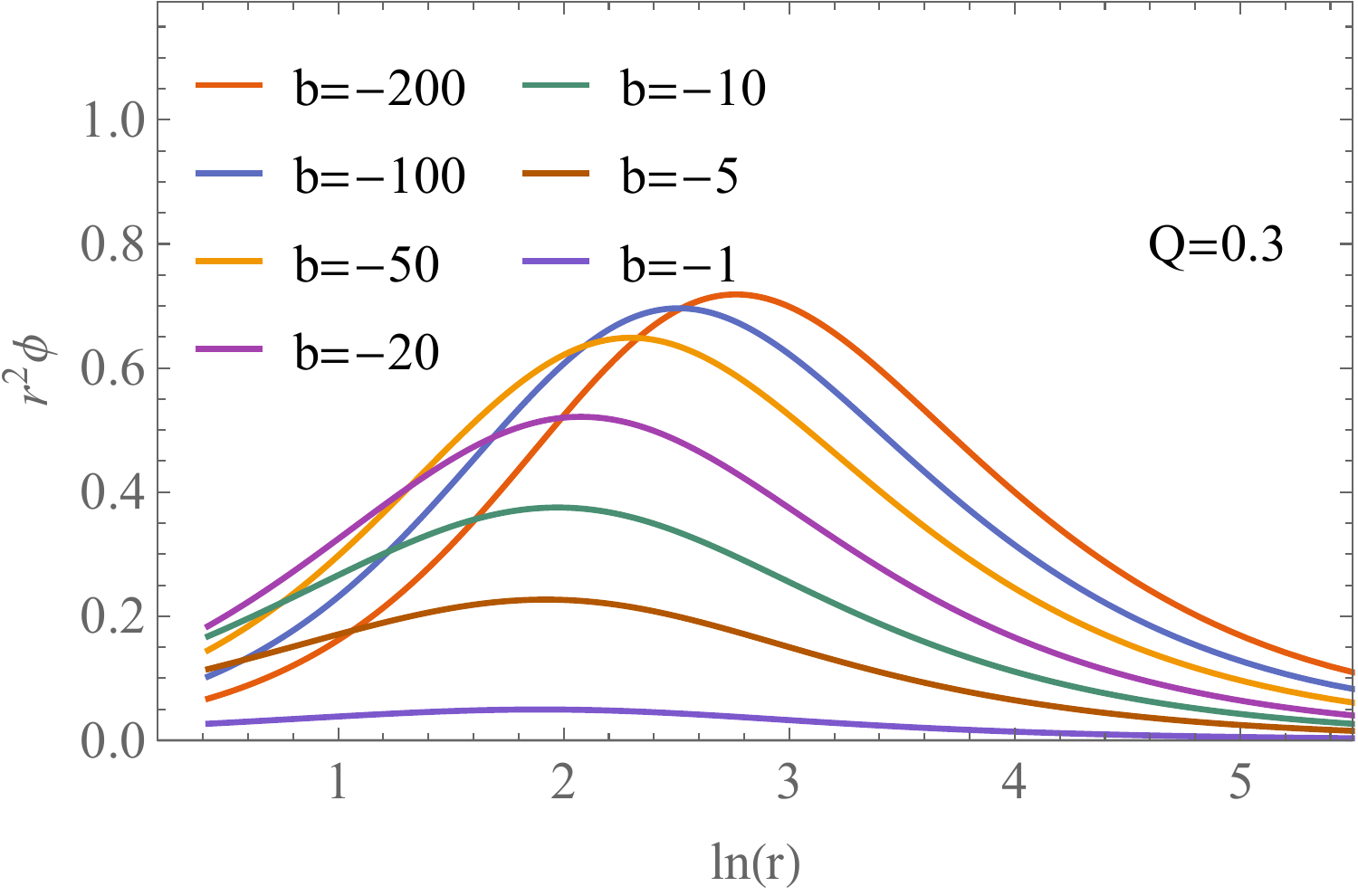}}\tabularnewline
\end{tabular}{\footnotesize\par}
\par\end{centering}
{\footnotesize{}\caption{\label{fig:Q93MSscalarProfile}{\small{} }The Misner-Sharp mass (upper)
and rescaled scalar profile $r^{2}\phi$ (lower) of the final hairy
black holes when $Q=0.9$ (left) and $0.3$ (right) for various $b$.
The dashed lines in the upper panels correspond to the Misner-Sharp
mass of the RN-AdS black hole with $Q=0.9$ and $Q=0.3$, respectively. Here $\Lambda=-0.03$.}
}{\footnotesize\par}
\end{figure*}

We show the Misner-Sharp mass and the scalar profile at late equilibrium time in the upper panels of Fig.\ref{fig:Q93MSscalarProfile}. The Misner-Sharp mass tends to the
ADM mass $M=1$ as the radius tends to infinity. For small values
of $-b$, the distribution of $M_{MS}$ is close to that of the corresponding
RN-AdS black hole. For large values of $-b$, $M_{MS}$ is constant
to a relatively large radial distance. At a larger radius, the radial dependence arises due to the presence of the scalar fields. We show the rescaled scalar field $r^{2}\phi$ which resembles the energy of the scalar in the spherical shell in the lower panels of Fig.\ref{fig:Q93MSscalarProfile}. The peak moves farther away from the black hole\footnote{Note that the scalar $\phi$ itself decreases with $r$ monotonically.} for larger $-b$.
  The reason can be deduced from the perturbation of \eqref{eq:phi}:
  \begin{equation}\label{eq:variation}
    \nabla^\mu \nabla_\mu \delta \phi = V\delta \phi,
  \end{equation}
  with the effective potential $V\equiv b^2 e^{-b\phi} F^{\mu\nu} F_{\mu\nu}$.
  Since $\phi$ decreases along $r$ direction, $e^{-b\phi}$ will be a rapidly decreasing function along $r$, forming a steep potential near the horizon. When increasing $-b$, the effective potential becomes steeper and drives the peak of  $r^2\delta \phi$ away from the horizon.
Combining our analysis for Fig. \ref{fig:BHsingular}, we conclude that the non-minimal coupling between the scalar and the Maxwell field plays two competing roles. On the one hand, it transfers the energy of the Maxwell field to the scalar. The scalar grows and is trapped by the black hole. On the other hand, it behaves as an effective repulsive force that drives the scalar away from the black hole. This is clear when one compares the left of Fig.\ref{fig:Q963b2010phi3f} and bottom of Fig.\ref{fig:Q93MSscalarProfile}. The $\phi_f$ at the infinity increases monotonically with $-b$, indicating that more energy is transferred to the scalar. But the scalar value at the black hole horizon increases at first and then decreases with $-b$ due to the repulsive effect of the non-minimal coupling.

{\footnotesize{}}
{\footnotesize\par}

\begin{figure*}
\begin{centering}
{\footnotesize{}}%
\begin{tabular}{cc}
{\footnotesize{}\includegraphics[width=0.41\textwidth]{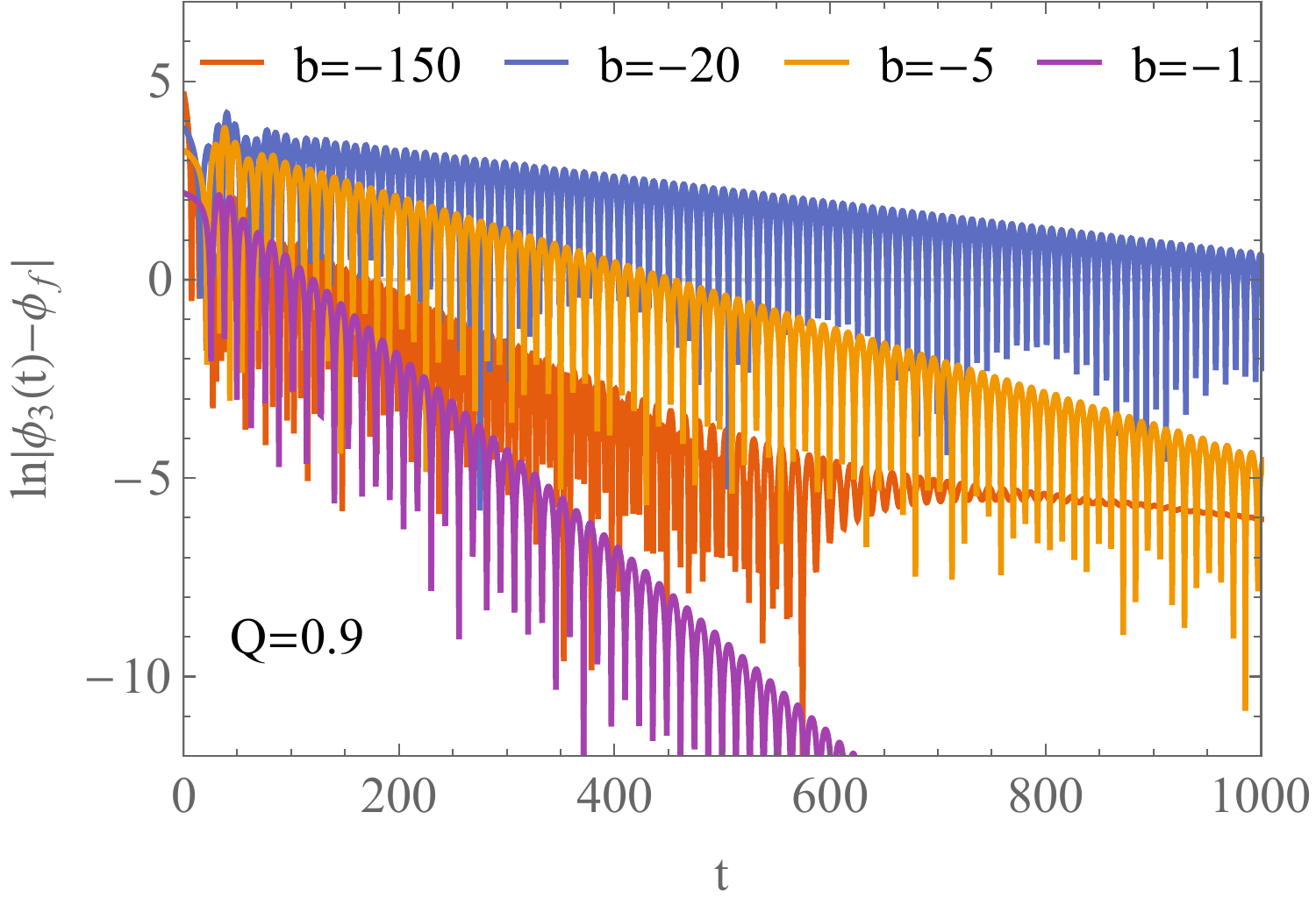}} & {\footnotesize{}\includegraphics[width=0.41\textwidth]{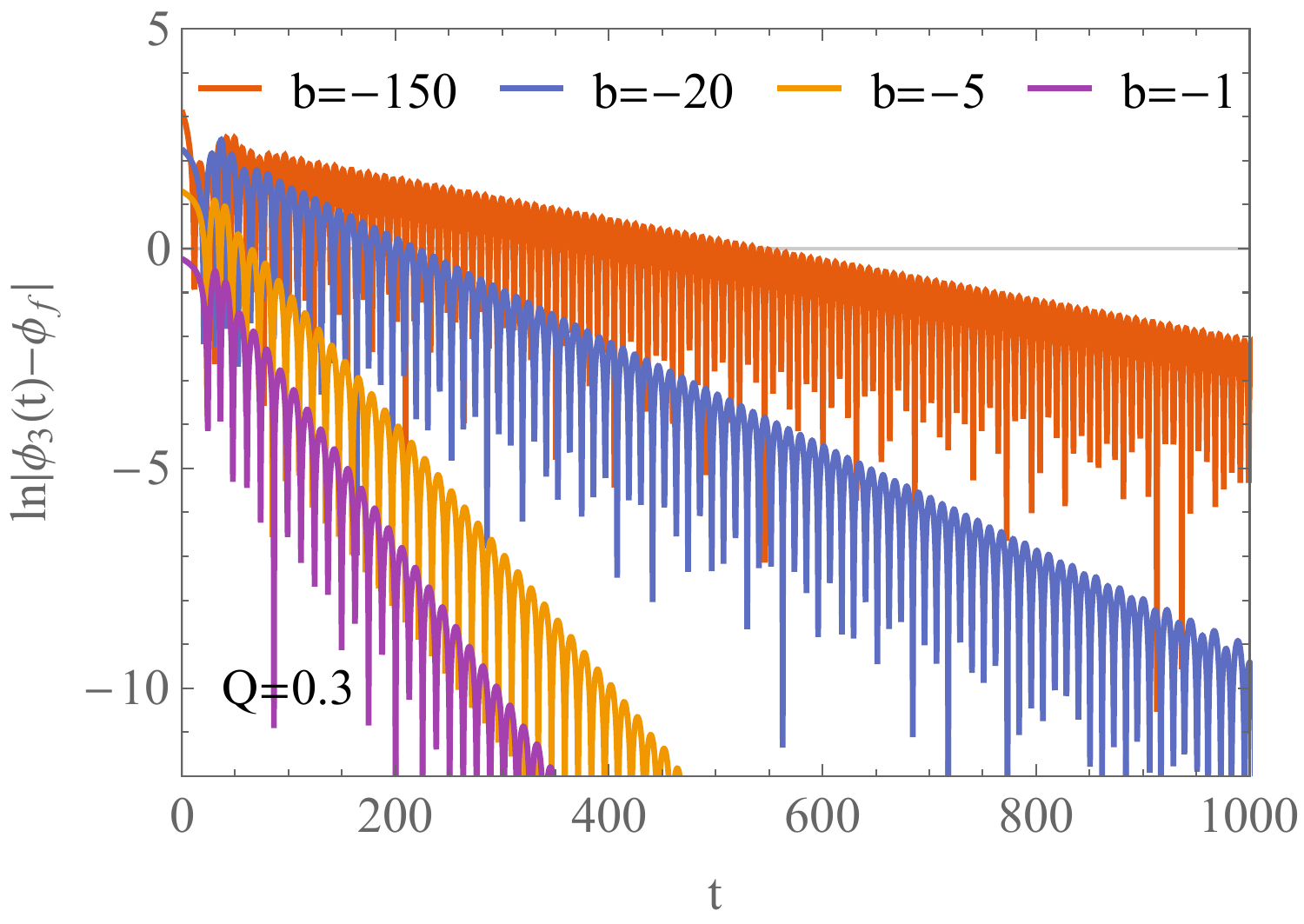}}\tabularnewline
{\footnotesize{}\includegraphics[width=0.41\textwidth]{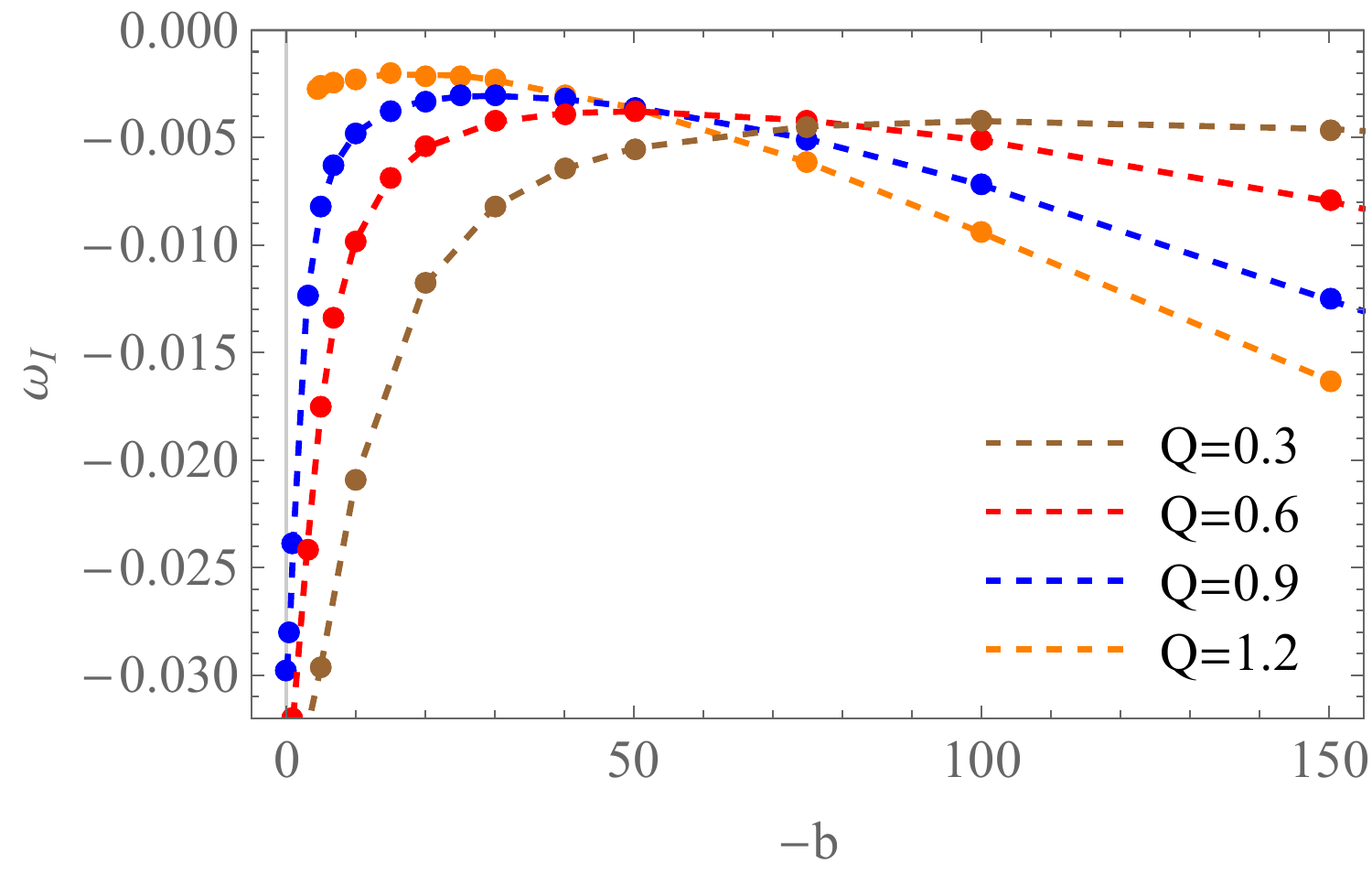}} & {\footnotesize{}\includegraphics[width=0.41\textwidth]{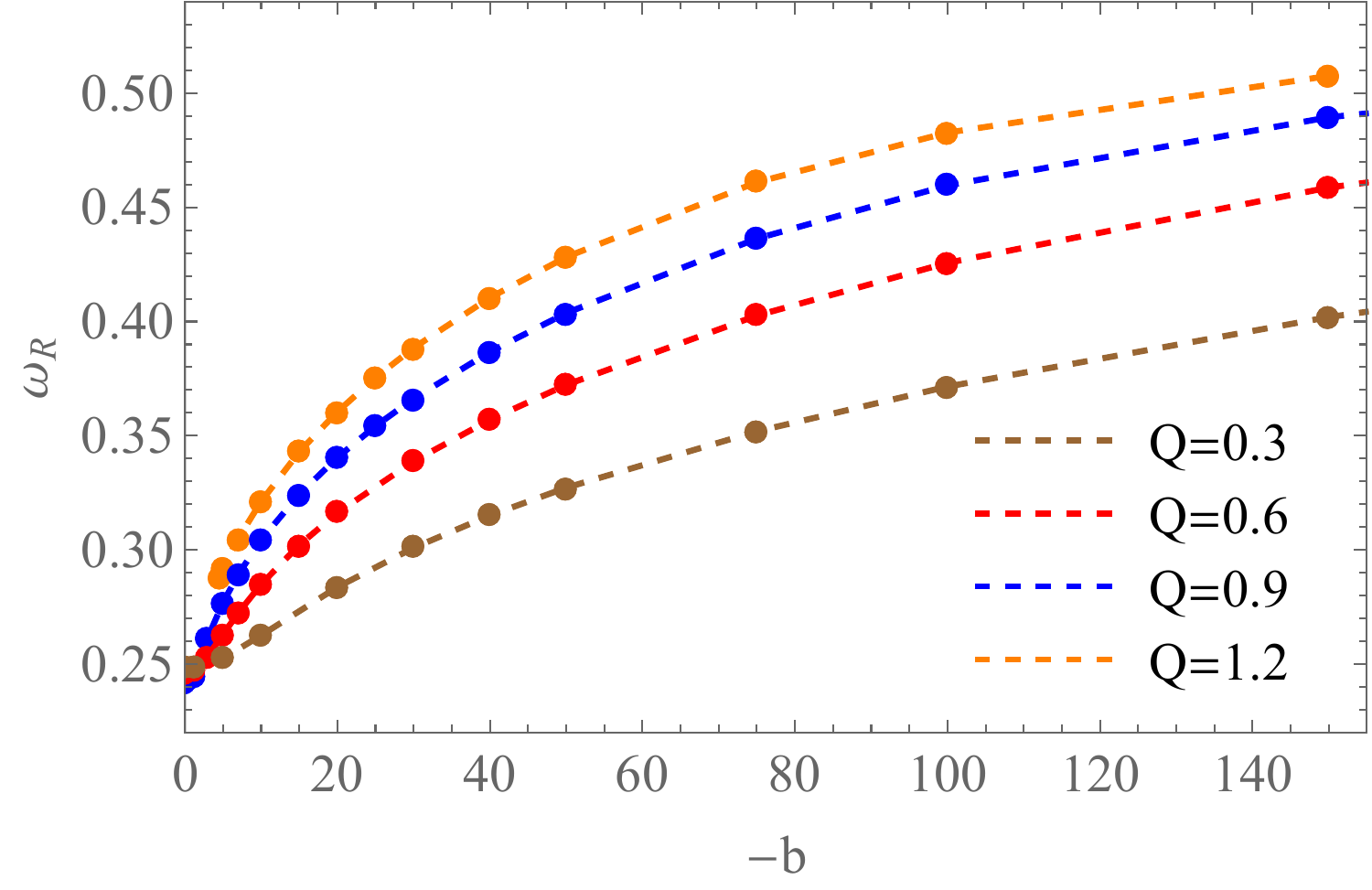}}\tabularnewline
\end{tabular}{\footnotesize\par}
\par\end{centering}
{\footnotesize{}\caption{\label{fig:Q963phi3wIR}{\small{} }The evolution of $\phi_{3}$ (upper)
and the dominant damping modes of $\phi_{3}$ v.s. $b$ (lower).  Here $\Lambda=-0.03$.}
}{\footnotesize\par}
\end{figure*}

Now we show the evolution of $\ln|\phi_{3}-\phi_{f}|$ in the upper panels
of Fig.\ref{fig:Q963phi3wIR}. It resembles the behavior of quasinormal
mode. The early time behavior of $\phi_{3}$ is closely related to
the initial perturbation. Then it oscillates with damping amplitude
and converges to the final value $\phi_{f}$ by a power-law.
Using Prony method \cite{Berti:2007dg}, we worked out the complex frequencies of the dominant damping modes shown in the lower panels of Fig.\ref{fig:Q963phi3wIR}.
The imaginary part $\omega_{I}$ increases at first and then decreases with $-b$,
indicating that the black hole with intermediate non-minimal coupling needs more time to settle down. This can be explained by the two competing roles $-b$ plays. When $-b$ is small, the energy is transferred from the Maxwell field to the scalar. The scalar is caught by the black hole and the initial RN-AdS black hole is destroyed, gradually developing into a hairy black hole solution. But now, the repulsive effect is weak such that the system takes a shorter time to settle down. As $-b$ increases, the initial RN-AdS black hole is also destroyed, but the repulsive effect becomes strong. The competition between the gravitational attraction and the repulsive effect makes the system needs more time to settle down. As $-b$ increases further, the repulsive effect dominates and makes the system settles down more easily. The real part $\omega_{R}$ increases monotonically with $-b$, indicating that the energy transfers faster to the scalar with stronger non-minimal coupling.

{\footnotesize{}}
{\footnotesize\par}

\begin{figure*}
\begin{centering}
{\footnotesize{}}%
\begin{tabular}{ccc}
{\footnotesize{}\includegraphics[width=0.32\textwidth]{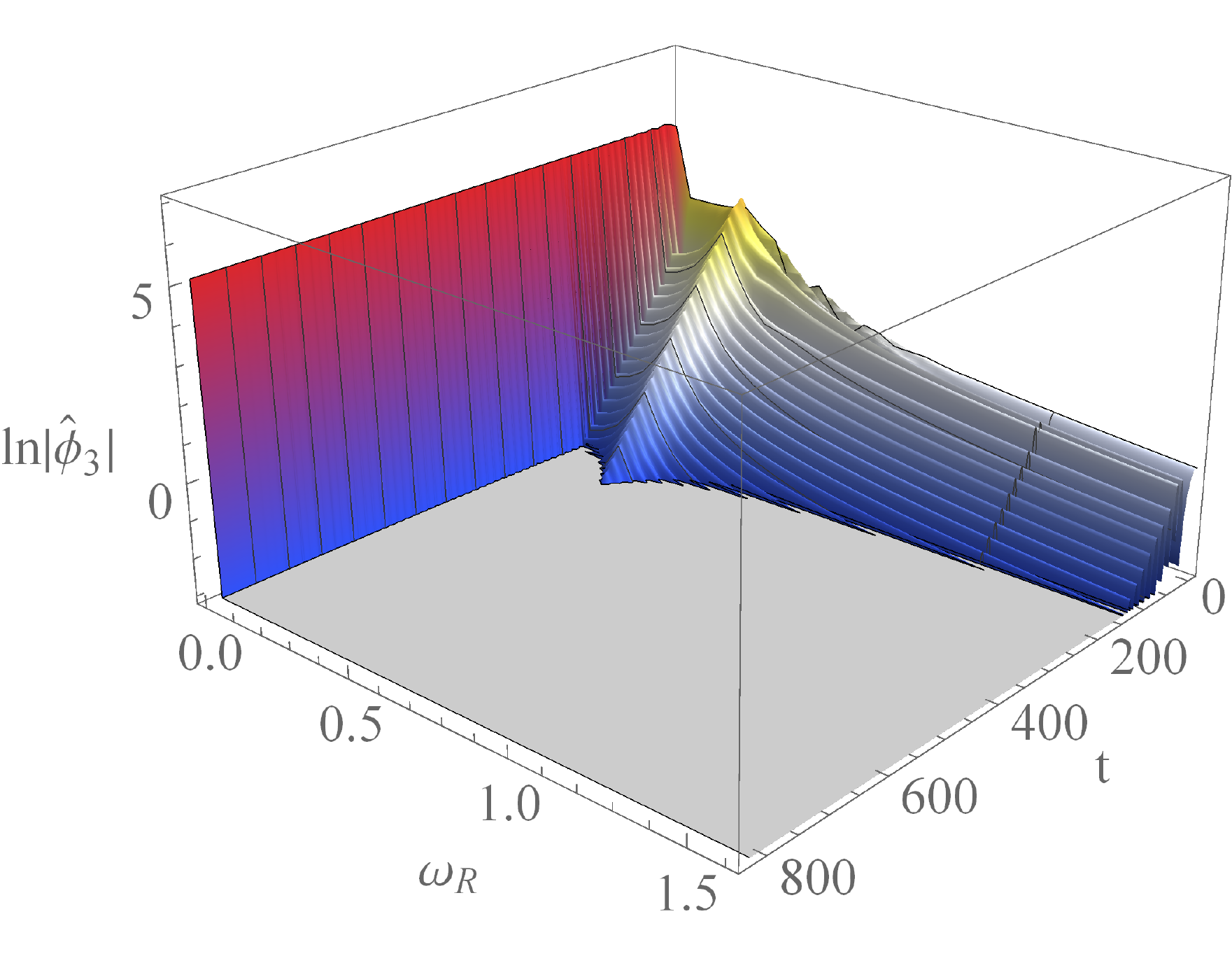}} & {\footnotesize{}\includegraphics[width=0.32\textwidth]{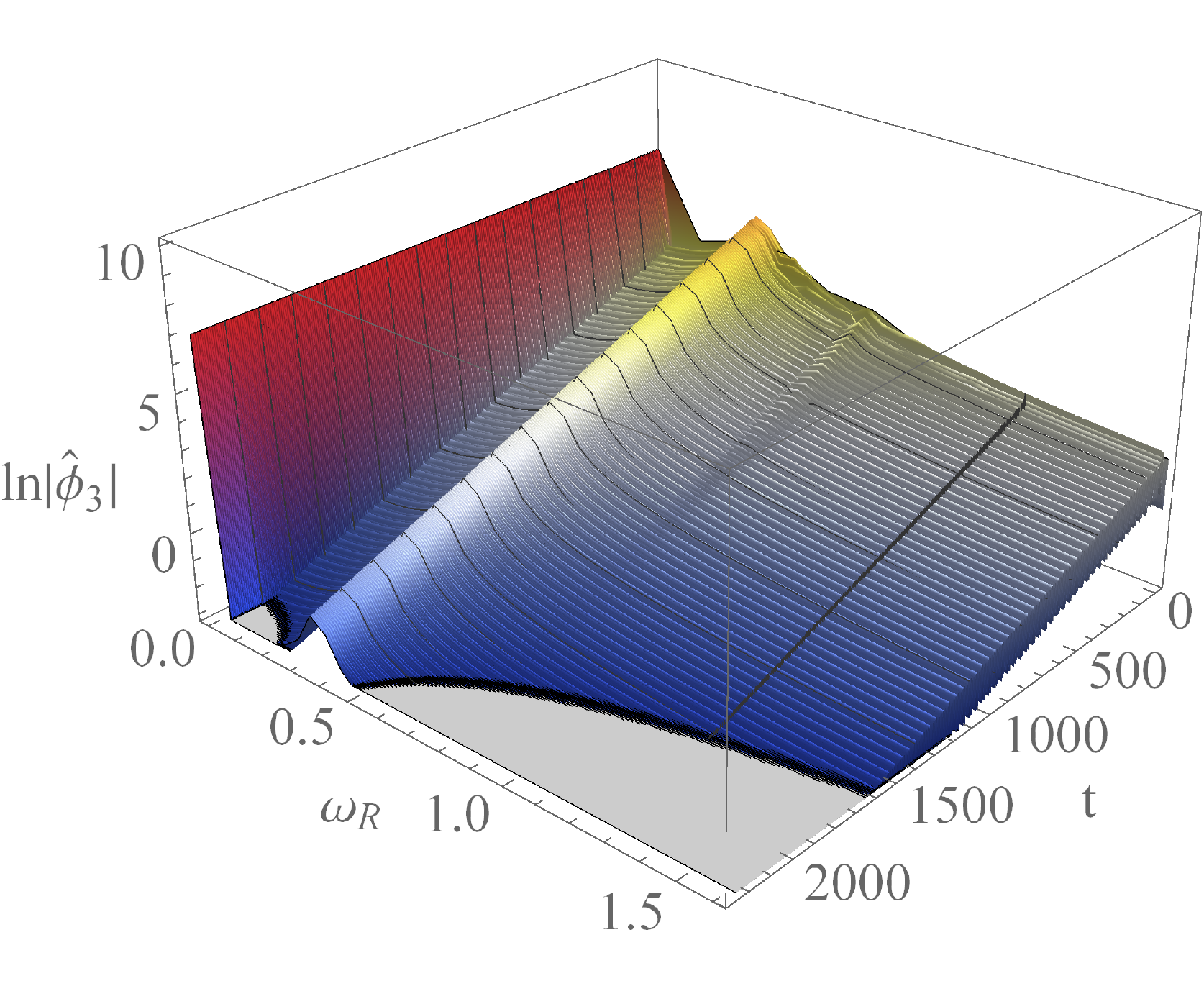}} & {\footnotesize{}\includegraphics[width=0.32\textwidth]{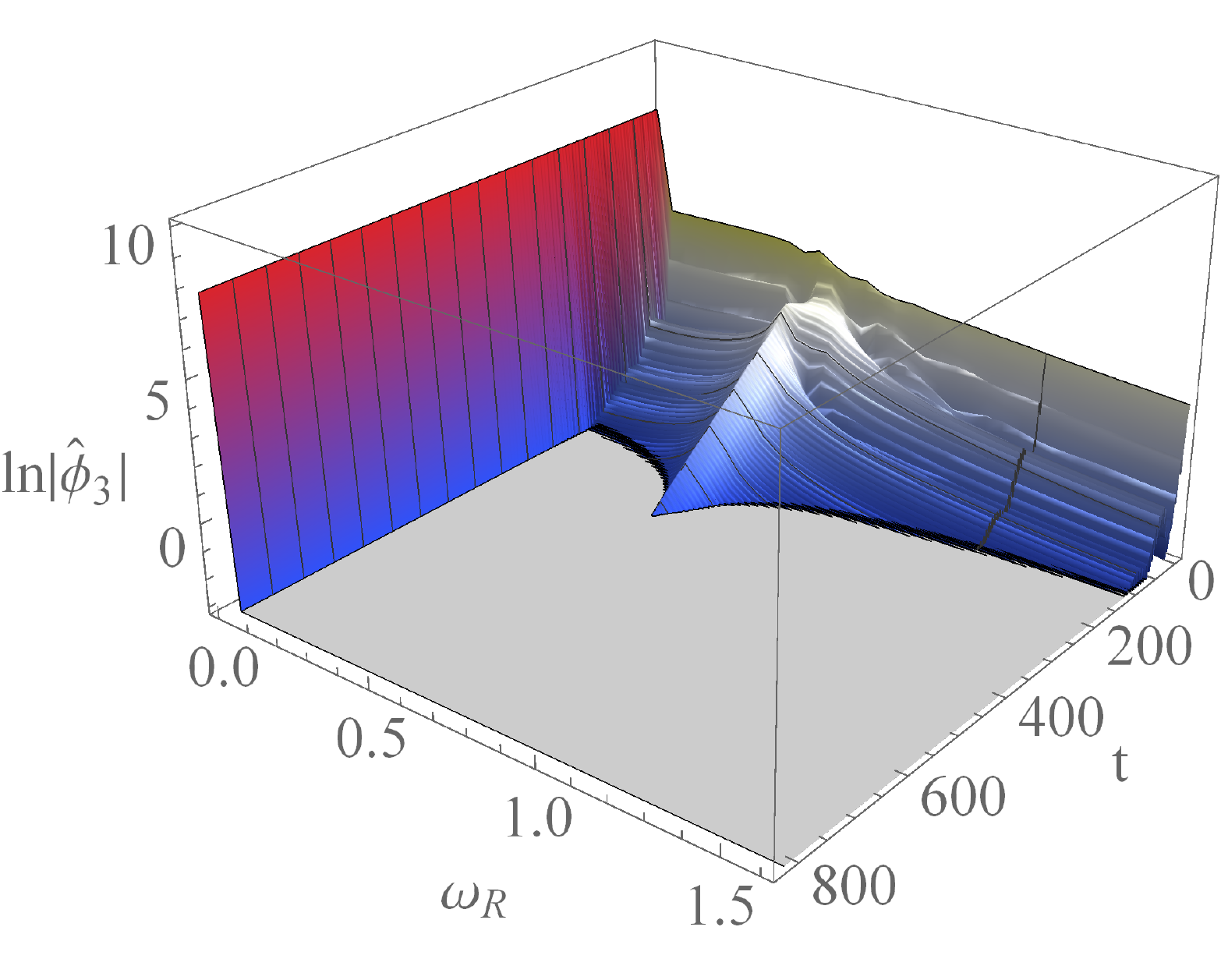}}\tabularnewline
\end{tabular}{\footnotesize\par}
\par\end{centering}
{\footnotesize{\caption{\label{fig:Q9phi3Fourier}{\small{} }The evolution of the logarithm
of amplitude $\ln|\hat{\phi}_{3}(t)|$ of discrete Fourier transformation
of $\phi_{3}(t)$ for $b=-1$ (left), $b=-30$ (middle) and $b=-200$
(right) when $Q=0.9$.  Here $\Lambda=-0.03$.}}
}
\end{figure*}

To more vividly demonstrate how each component modes evolve, we show the evolution of the amplitudes of each component
modes in Fig.\ref{fig:Q9phi3Fourier}. These are calculated by partitioning
the time axis into many overlapping subintervals with an appropriate
offset and performing discrete Fourier transformation on each of these subintervals \cite{Bosch:2016vcp}. We see that all modes decay at late times except the zero mode. The oscillating frequency and decaying rate of the dominant decaying mode are consistent with the $\omega_{R}$ and $\omega_{I}$ from Prony method. From the left to the right in Fig.\ref{fig:Q9phi3Fourier} are cases for $-b = 1, 30, 200$, respectively. The real part of the frequency of the dominant decaying mode increases for larger $-b$ and the energy of the scalar increases faster as $-b$ increases. This is consistent with our analysis for the bottom right of Fig.\ref{fig:Q963phi3wIR}. We also find that the dominant decaying mode takes a longer time to decay at the intermediate value of $-b$. This is in accordance with the convex behavior of $\omega_I$ with $-b$.

{\footnotesize{}}
{\footnotesize\par}

\begin{figure*}
\begin{centering}
{\footnotesize{}}%
\begin{tabular}{ccc}
{\footnotesize{}\includegraphics[width=0.32\textwidth]{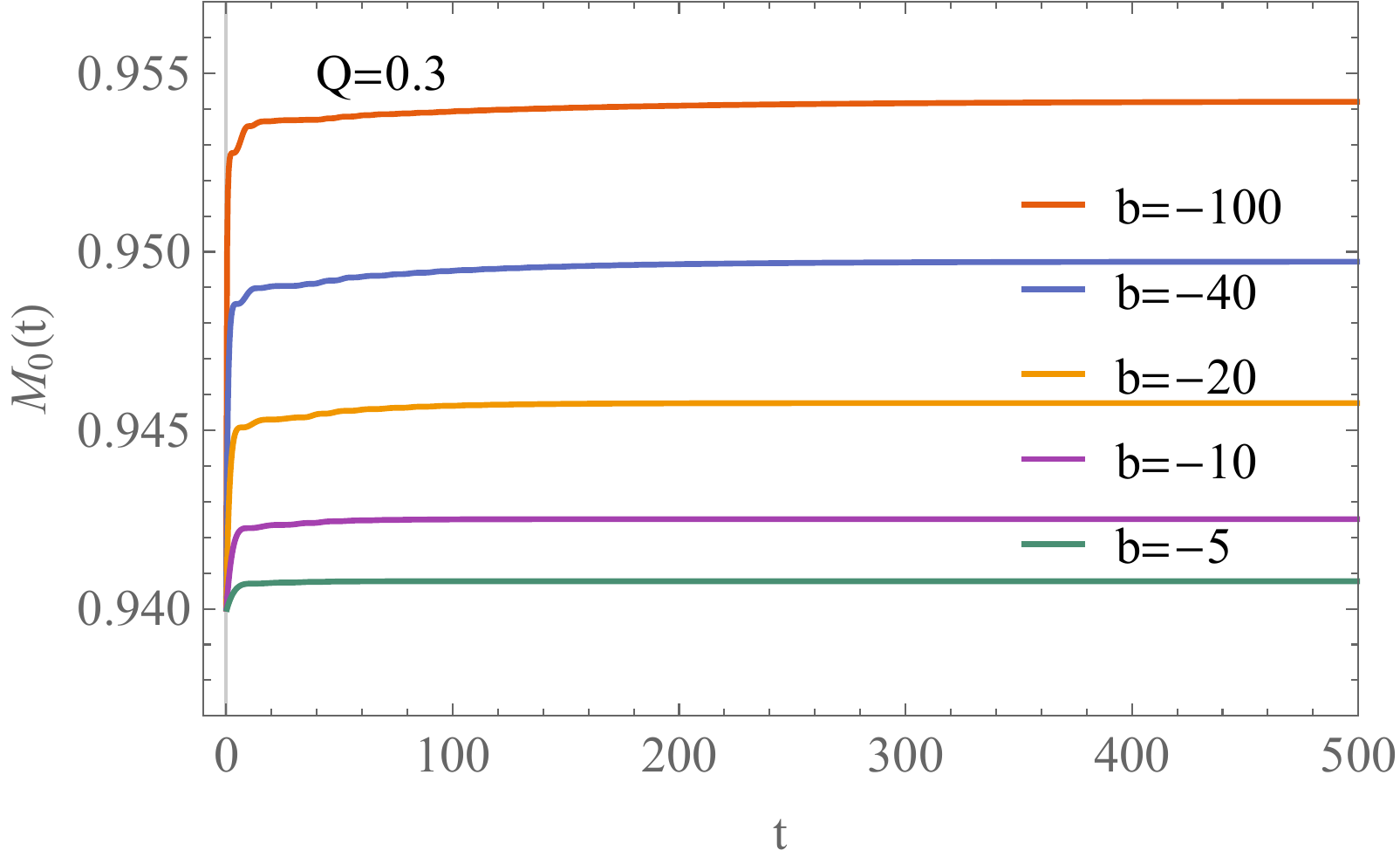}} & {\footnotesize{}\includegraphics[width=0.32\textwidth]{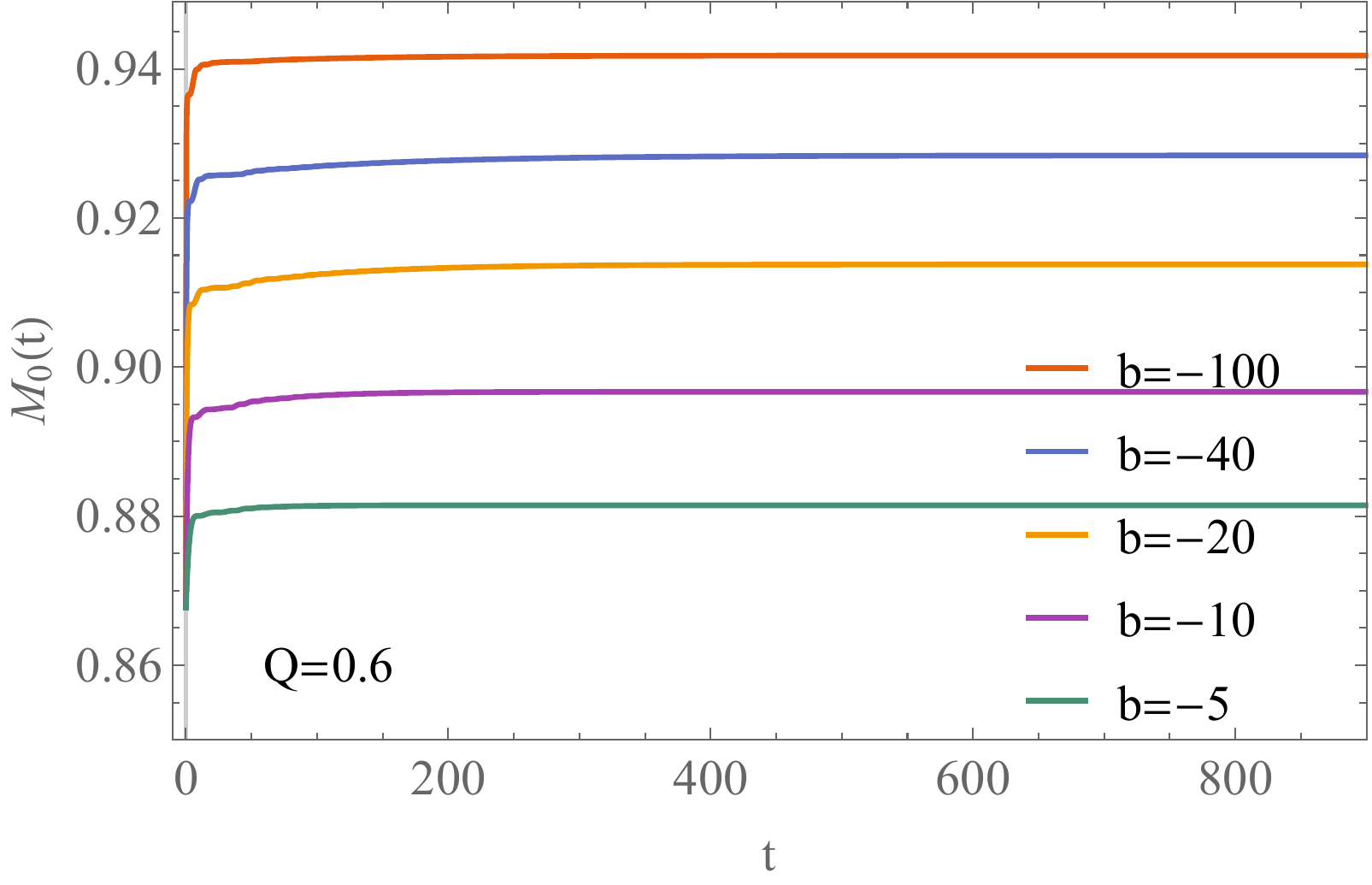}} & {\footnotesize{}\includegraphics[width=0.32\textwidth]{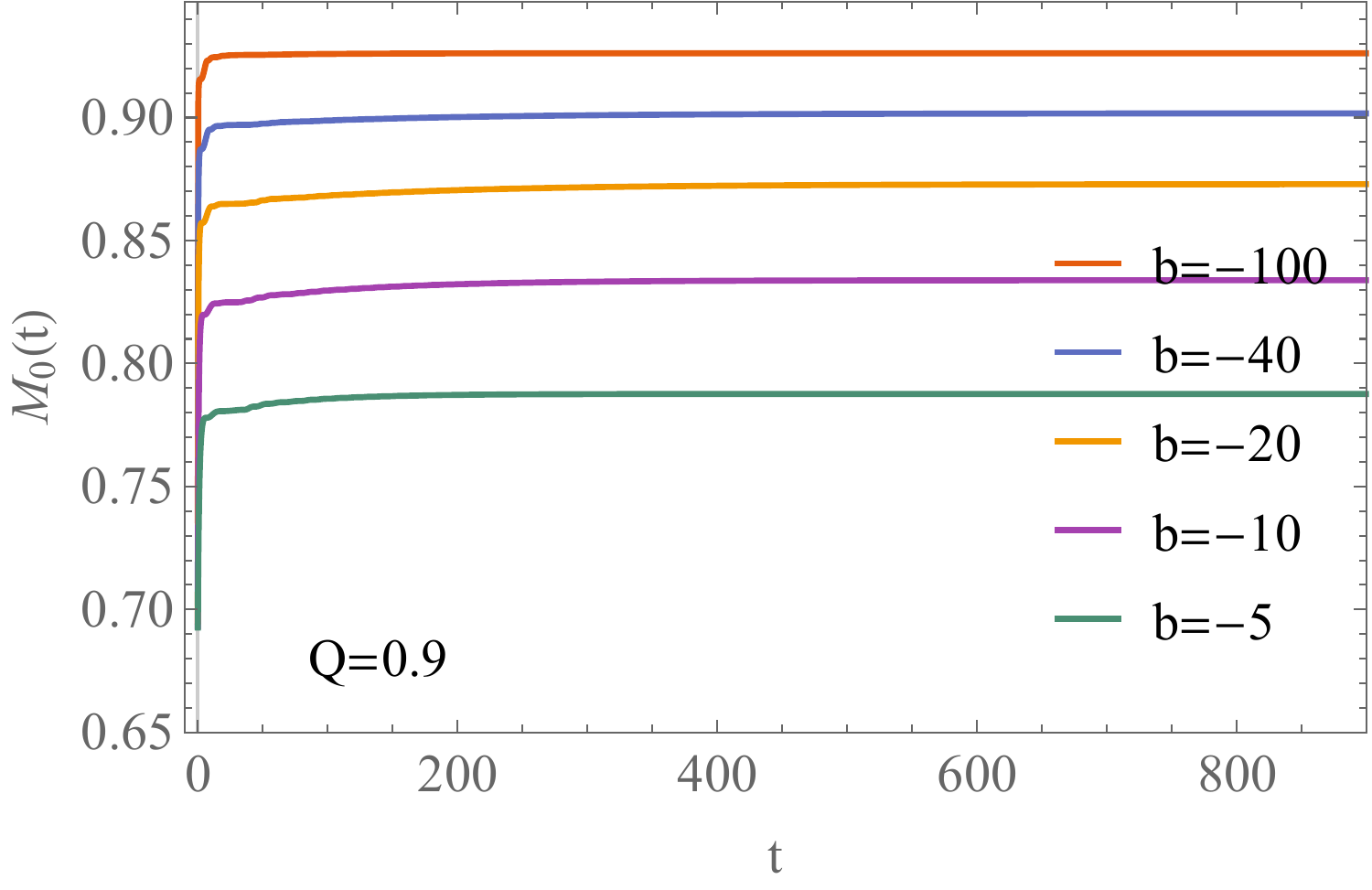}}\tabularnewline
{\footnotesize{}\includegraphics[width=0.32\textwidth]{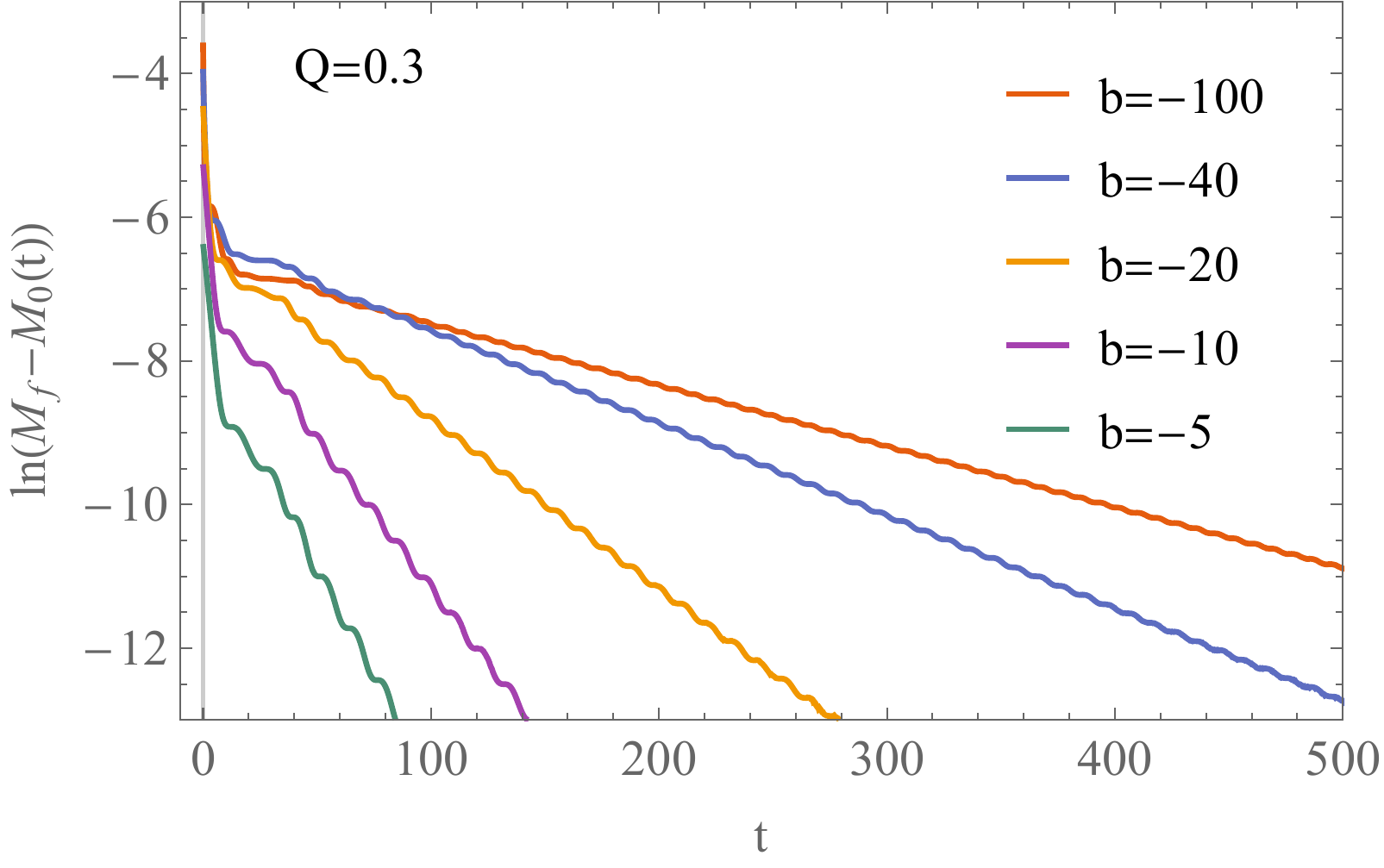}} & {\footnotesize{}\includegraphics[width=0.32\textwidth]{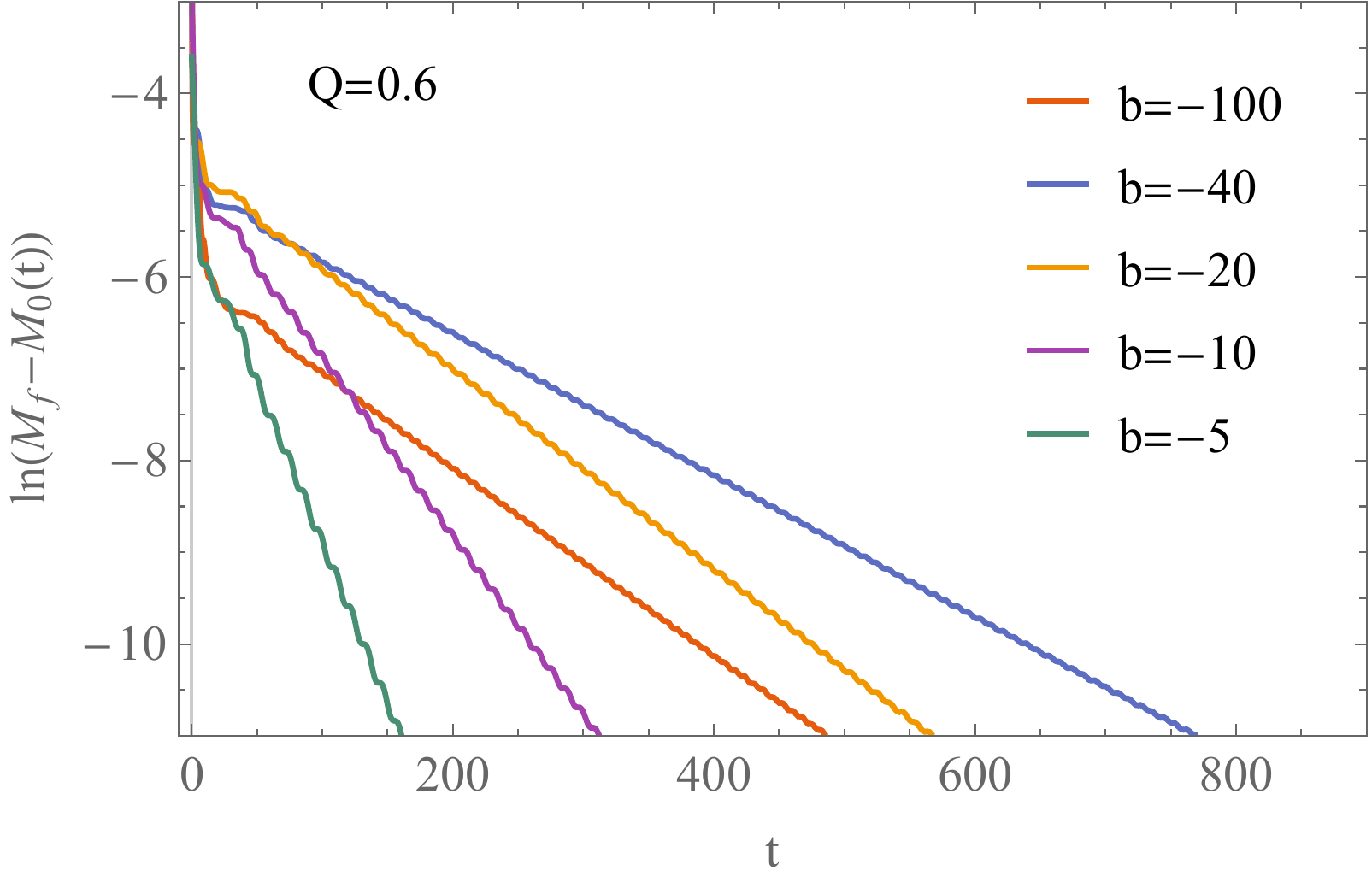}} & {\footnotesize{}\includegraphics[width=0.32\textwidth]{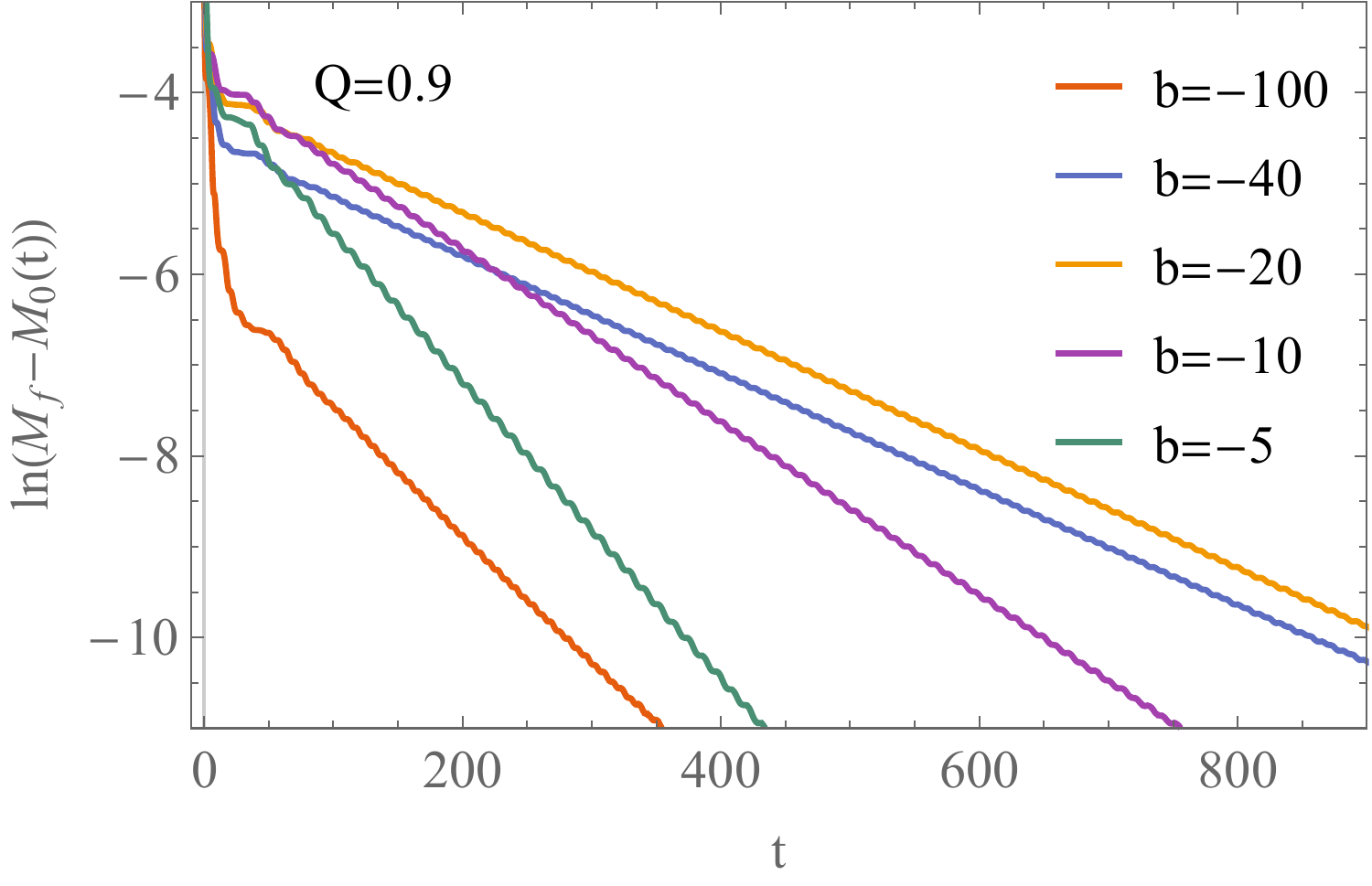}}\tabularnewline
\end{tabular}{\footnotesize\par}
\par\end{centering}
{\footnotesize{}\caption{\label{fig:Q963irrM}{\small{} }The evolution of irreducible mass
$M_{0}$ (upper panels) and $\ln(M_{f}-M_{0})$ of the black hole
for various $b$ when $Q=0.3,0.6$ and $0.9$. Here $M_{f}$ is the final value of the irreducible mass of the black hole.  Here $\Lambda=-0.03$.}
}{\footnotesize\par}
\end{figure*}

Now, we study the evolution of the irreducible mass of the black hole,
which is shown in Fig.\ref{fig:Q963irrM}. At early times, the irreducible mass increases abruptly. This is different from the case in EMS model where the irreducible mass changes little at early times \cite{Zhang:2021etr}. RN-AdS black hole solves the EMS theory, the instability of this solution triggered by the scalar field is local and it costs time for the scalar perturbation traveling from the initial position to the horizon. Only when the scalar perturbation arrives at the horizon does the black hole begin to grow.
However, for the EMD theory, the RN-AdS black hole metric does not solve the model, from the very beginning of the process of evolution the energy outside the horizon is redistributed everywhere, and some of it is swallowed by the central black hole.

{\footnotesize{}}
{\footnotesize\par}

\begin{figure*}
\begin{centering}
{\footnotesize{}}%
\begin{tabular}{cc}
{\footnotesize{}\includegraphics[width=0.41\textwidth]{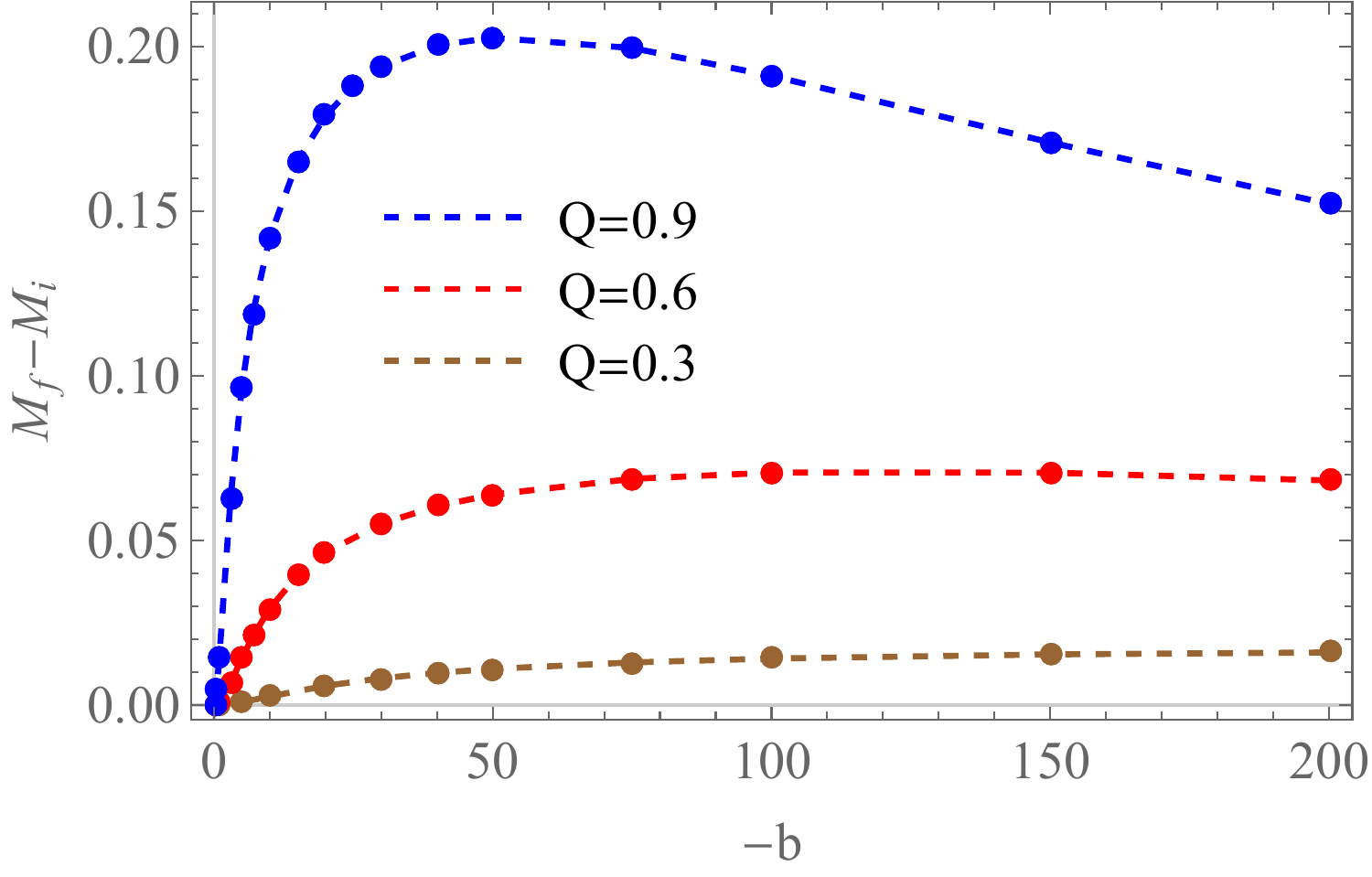}} &
 {\footnotesize{}\includegraphics[width=0.41\textwidth]{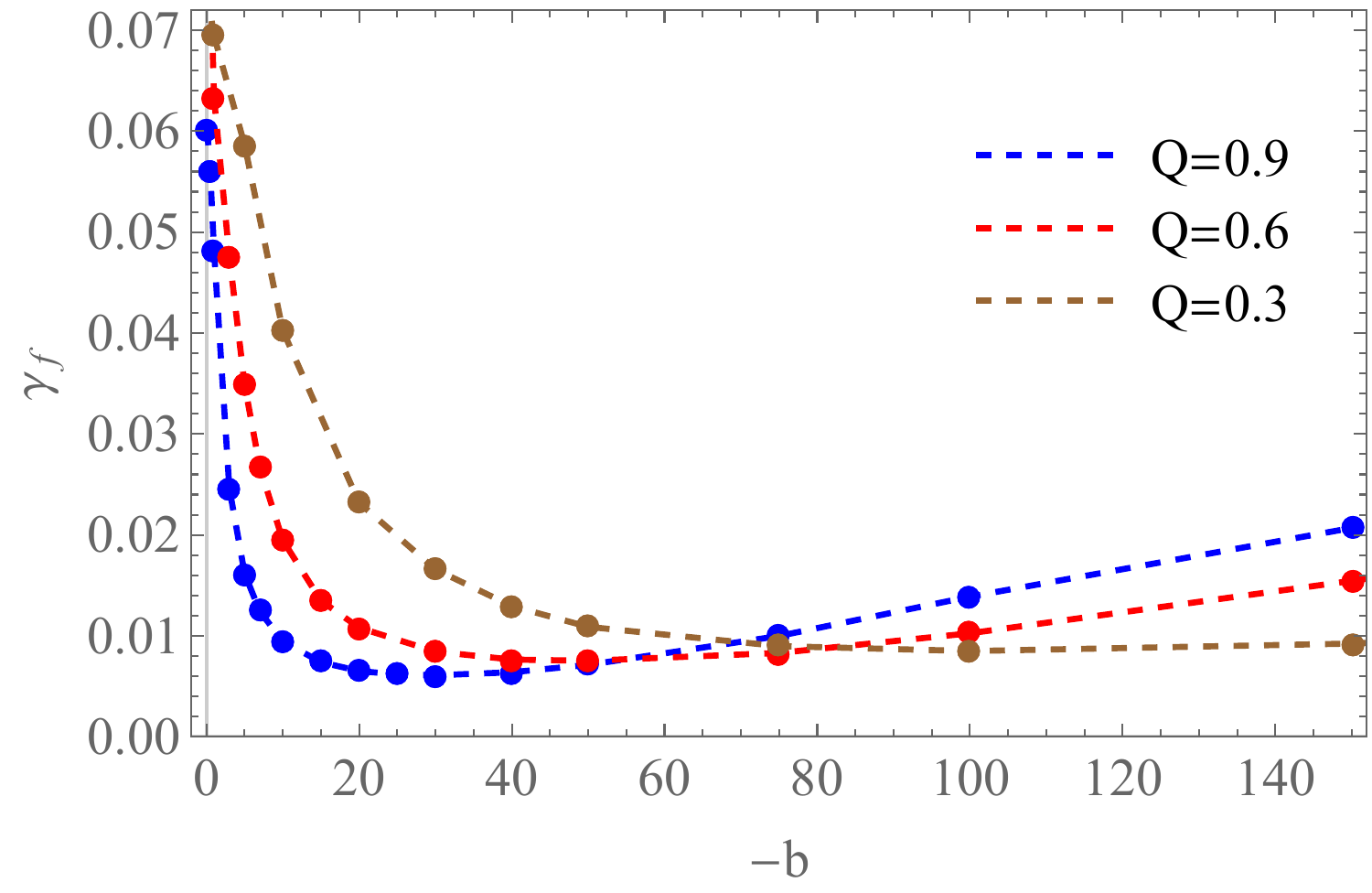}}\tabularnewline
\end{tabular}{\footnotesize\par}
\par\end{centering}
{\footnotesize{}\caption{\label{fig:Q963irrMfindex}{\small{} }The increment of the black
hole irreducible mass $M_{f}-M_{i}$ (left) and saturating rate $\gamma_{f}$
for various $b$ when $Q$ is fixed.  Here $\Lambda=-0.03$.}
}{\footnotesize\par}
\end{figure*}

At late times, the irreducible mass saturates to the final value $M_{f}$ with behavior
\begin{align}
M_{0}(t)\simeq & M_{f}-e{}^{-\gamma_{f}t+c_{f}}.\label{eq:MirrExp}
\end{align}
This is similar to the case in the EMS model at late times. $\gamma_{f}$ is the saturating rate and depends on $-b$. Note that the irreducible mass does not decrease here.  The increment of the black hole irreducible mass $M_{f}-M_{i}$ is shown in the left of Fig.\ref{fig:Q963irrMfindex}. Here $M_{i}$ is the initial irreducible mass of the black hole. When $Q$ is small, $M_{f}-M_{i}$ increases monotonically as a function of $-b$. When $Q$ becomes large, $M_{f}-M_{i}$ is no longer a monotonic increasing function of $-b$. Instead, it increases at first and then decreases as $-b$ increases. For the model with a large $Q$, the strength of the effective repulsive force from the Maxwell field affecting on the scalar field increase fast and balance the black hole's attractive force thus suppressing the energy flux from outside to inside of the horizon.

{\footnotesize{}}
{\footnotesize\par}
The saturating rate $\gamma_{f}$ is shown in  Fig.\ref{fig:Q963irrMfindex}. When $Q$ is small, $\gamma_{f}$ decreases almost monotonically with
$-b$. When $Q$ is large, $\gamma_{f}$ decreases at first and then increases with $-b$. The irreducible mass saturates faster when $-b$ is small or large enough. This is consistent with Fig.\ref{fig:Q963phi3wIR}.
An interesting fact  we find is that at late times, there is
  \begin{equation}\label{eq:gammaomega}
    \gamma_{f} = -2\omega_{I},
  \end{equation}
which can be obtained by comparing the $\gamma_{f}$ in Fig.\ref{fig:Q963irrMfindex} and $\omega_{I}$ in Fig. \ref{fig:Q963phi3wIR}. In fact, the irreducible mass is nothing but the value of $\zeta$ at the horizon. Hence the late time evolution of the irreducible mass can be deduced from the evolution of $\zeta$. It is known that the perturbation of the scalar field invokes the back-reaction of the metric only at the second order \cite{Brito:2015oca}. Namely, there is
\begin{equation}\label{eq:secondorder}
  \delta \zeta \sim (\delta \phi)^2 + \cdots,
\end{equation}
where $\cdots$ represents other possible perturbations. Therefore, the saturating rate $-\gamma_f$ of the $\delta \zeta$ should be twice of $\omega_I$. This relation holds also in the EMS theory \cite{Zhang:2021etr}.

\subsection{Effects of charge $Q$ on the black hole evolution}\label{subsec:q}

In this subsection, we fix $\Lambda=0.03$ and choose certain $b$ to study the effects of charge $Q$ on the black hole evolution. The final value $\phi_f$ of $\phi_3$ is shown in the right panel of Fig.\ref{fig:Q963b2010phi3f}. There is always a hairy black hole solution when $Q$ is nonvanishing.
The scalar hair increases monotonically with both $Q$ and $-b$. Note that our numerical evolution codes crash for $Q>2,3,4$ when $-b=20,50,150$, respectively. This implies that the charge to mass ratio of the regular hairy black hole solution can not be too large \cite{Garfinkle:1990qj}.

{\footnotesize{}}
\begin{figure*}
\begin{centering}
{\footnotesize{}}%
\begin{tabular}{cc}
{\footnotesize{}\includegraphics[width=0.45\textwidth]{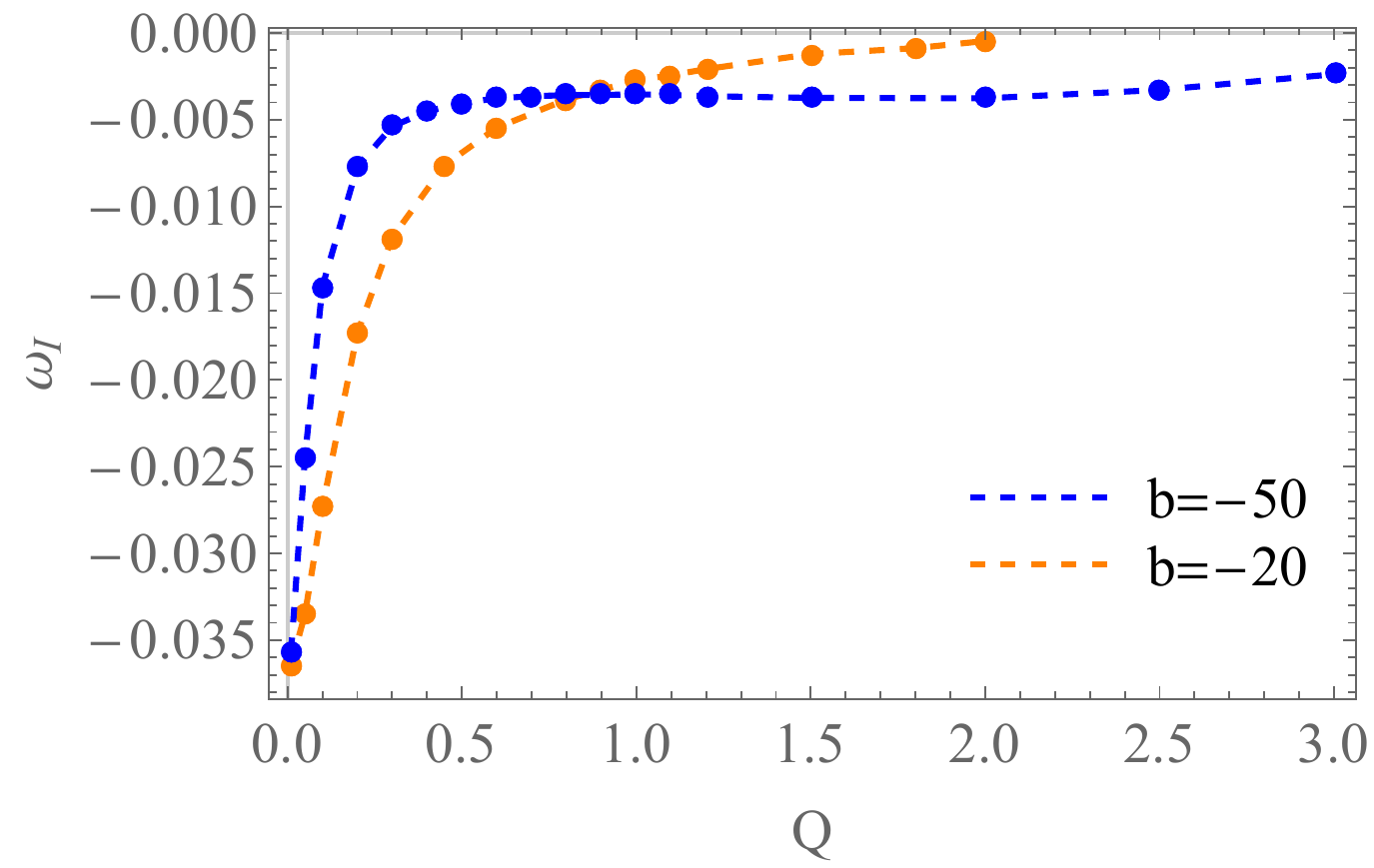}} & {\footnotesize{}\includegraphics[width=0.43\textwidth]{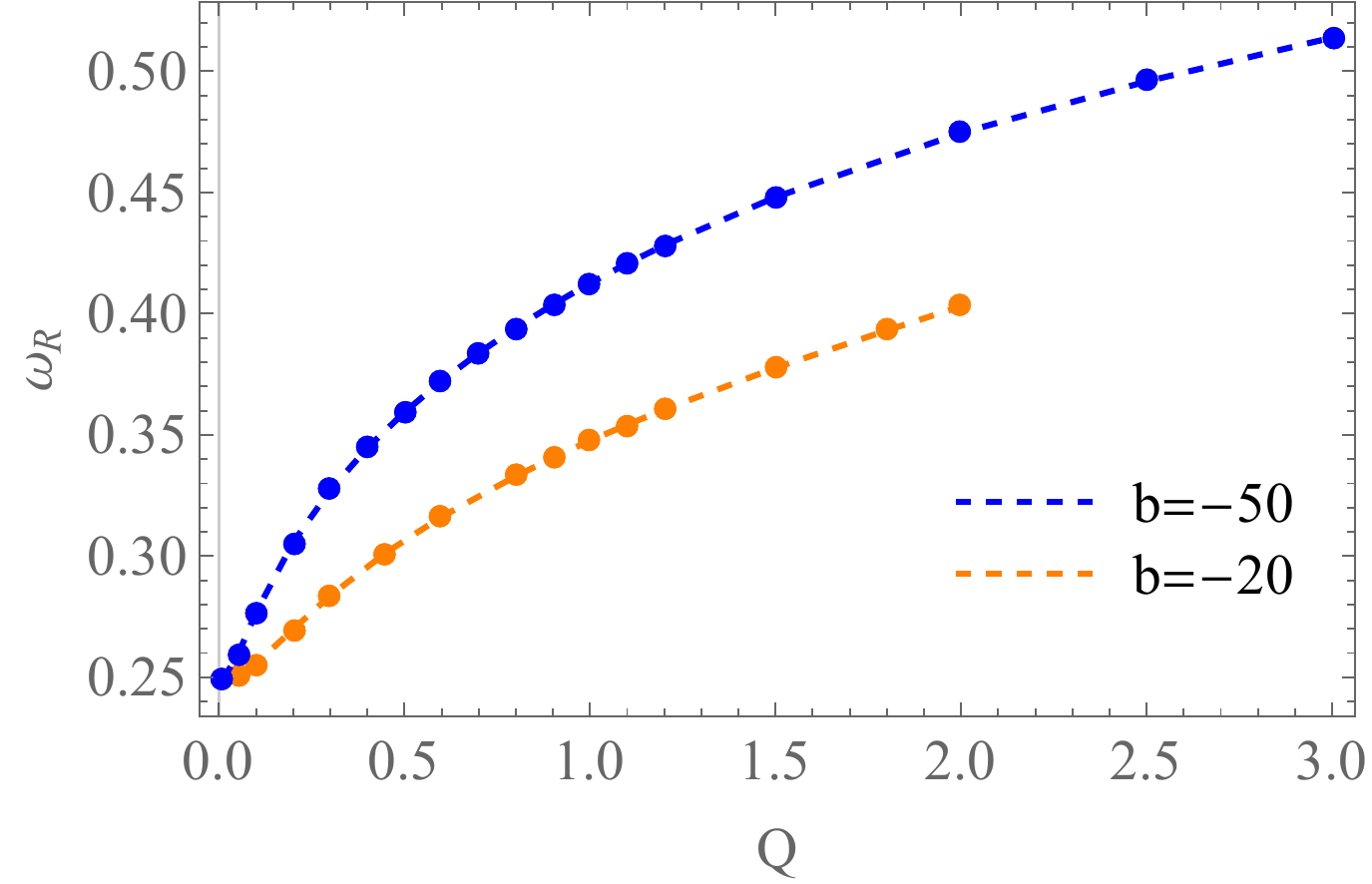}}\tabularnewline
\end{tabular}{\footnotesize\par}
\par\end{centering}
{\footnotesize{}\caption{\label{fig:b2010phi3wIR}{\small{} }The imaginary and real part of the frequencies of  the dominant damping modes of $\phi_{3}$ v.s. $Q$ when $b=-150,-50$
and $-20$.  Here $\Lambda=-0.03$.}
}{\footnotesize\par}
\end{figure*}
{\footnotesize\par}
The evolution of $\phi_3$ still resembles the behavior of the quasinormal mode. All the modes damp exponentially except the zero modes. The frequencies of the dominant modes are shown in Fig.\ref{fig:b2010phi3wIR}. The real part increases monotonically with $Q$ and $-b$. The imaginary part increases monotonically with $Q$ only when $-b$ is small. For large $-b$, it increases with small $Q$ at first, then it decreases with intermediate $Q$ and then increases with large $Q$.
In fact, when $-b$ is fixed, the strength of the non-minimal coupling is controlled by charge.  When $Q$ is small, the energy of the Maxwell field is transferred to the scalar but the repulsive effect is small, resulting in a hairy black hole solution in a relatively short time. As $Q$ increases, the repulsive effects of the non-minimal coupling become stronger and its competition between the gravitational attraction makes the system need more time to settle down.

\begin{figure*}
\begin{centering}
{\footnotesize{}}%
\begin{tabular}{cc}
 {\footnotesize{}\includegraphics[width=0.43\textwidth]{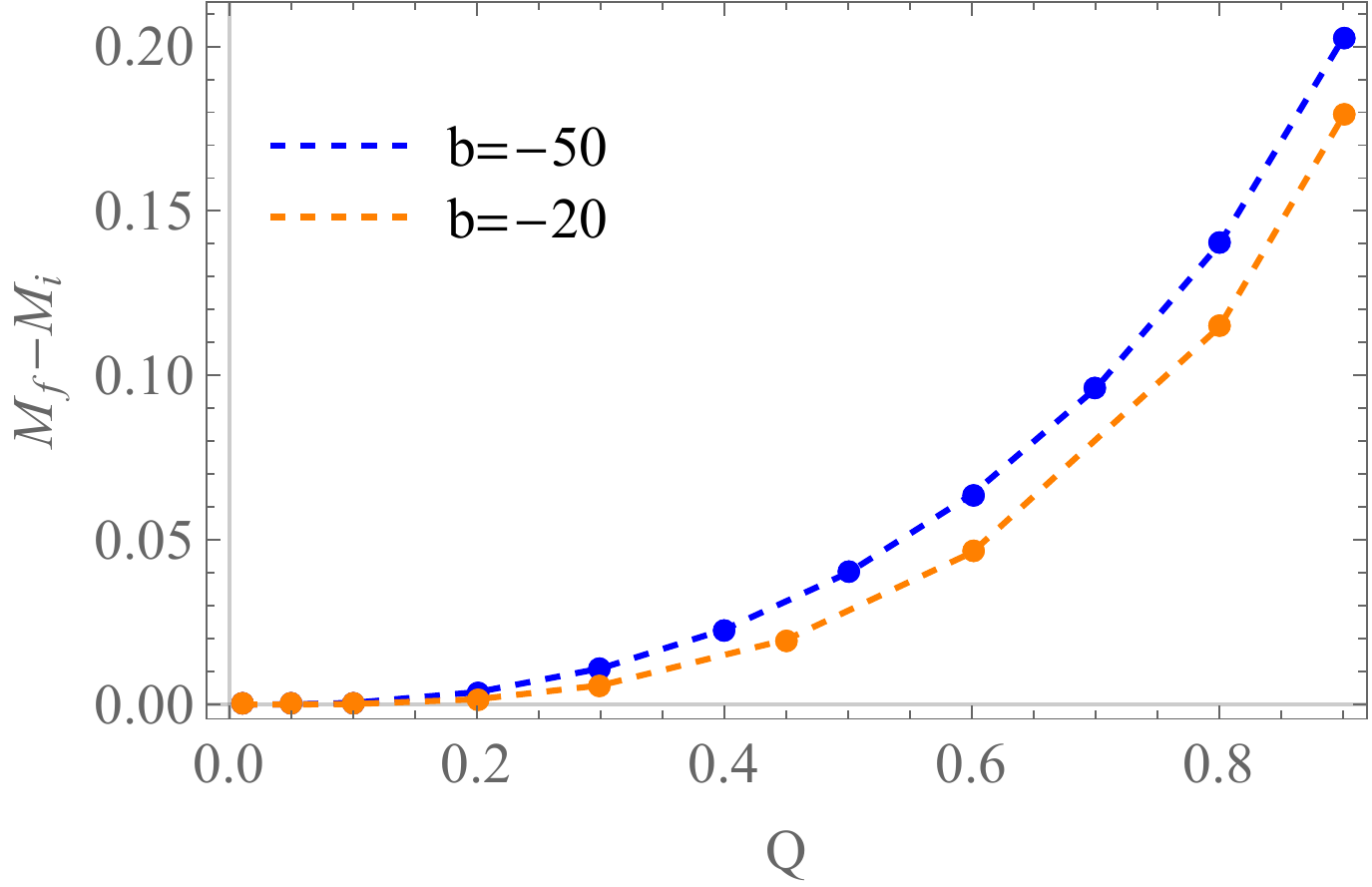}} & {\footnotesize{}\includegraphics[width=0.43\textwidth]{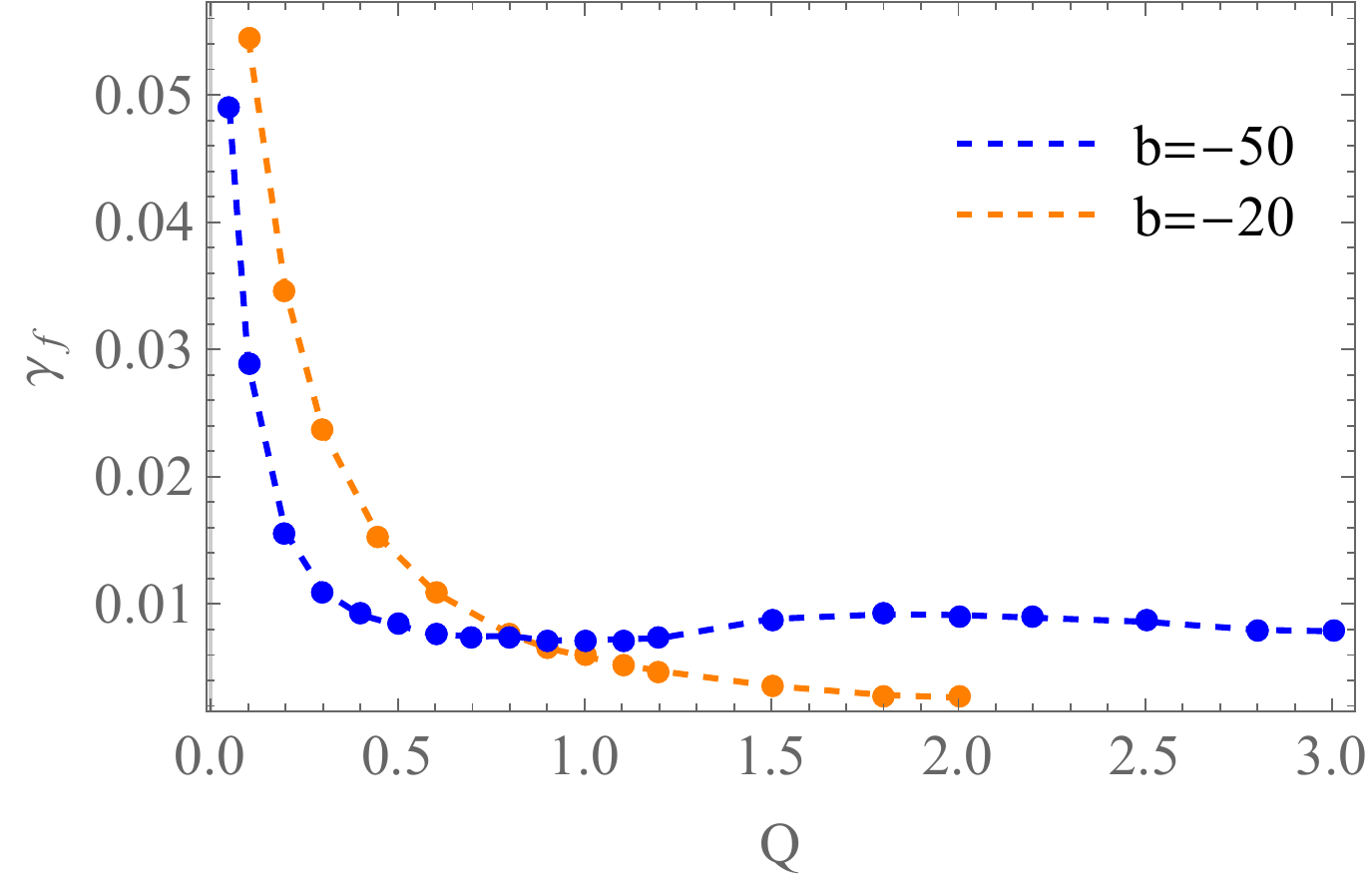}}\tabularnewline
\end{tabular}{\footnotesize\par}
\par\end{centering}
{\footnotesize{}\caption{\label{fig:b3020irrM}{\small{} }The increment of
irreducible mass of the black hole and the  saturating rate for various $Q$ when $b=-50$
 and $-20$.  Here $\Lambda=-0.03$.}
}{\footnotesize\par}
\end{figure*}

The irreducible mass $M_0$ of the black hole still increases abruptly at early times, and then saturates to the final value exponentially. The left panel of Fig.\ref{fig:b3020irrM} shows\footnote{We only show the increment when $Q<1$ since when $Q>1$, the initial configuration is a naked singularity and there is no corresponding $M_i$.} that the increment of the irreducible mass $M_f - M_i$ increases monotonically with $Q$. The right panel shows that the rate $\gamma_f$ decreases with $Q$ when $-b$ is small. For large $-b$, the rate decreases with small $Q$ at first and then increases with intermediate $Q$ and then decreases again with large $Q$ when $-b$ is large. When the coupling parameter $-b$ is small, it takes a shorter time to settle down for systems with small $Q$. This is consistent with the results from the imaginary part of the frequencies of the dominant modes in the left panel of Fig.\ref{fig:b2010phi3wIR} { and there is still $\gamma_f=-2\omega_I$ at late times.}

{\footnotesize{}}
{\footnotesize\par}

\subsection{Effects of the cosmological constant $\Lambda$ on the black hole
evolution}\label{subsec:l}

\begin{figure*}
\begin{centering}
{\footnotesize{}}%
\begin{tabular}{ccc}
{\footnotesize{}\includegraphics[width=0.31\textwidth]{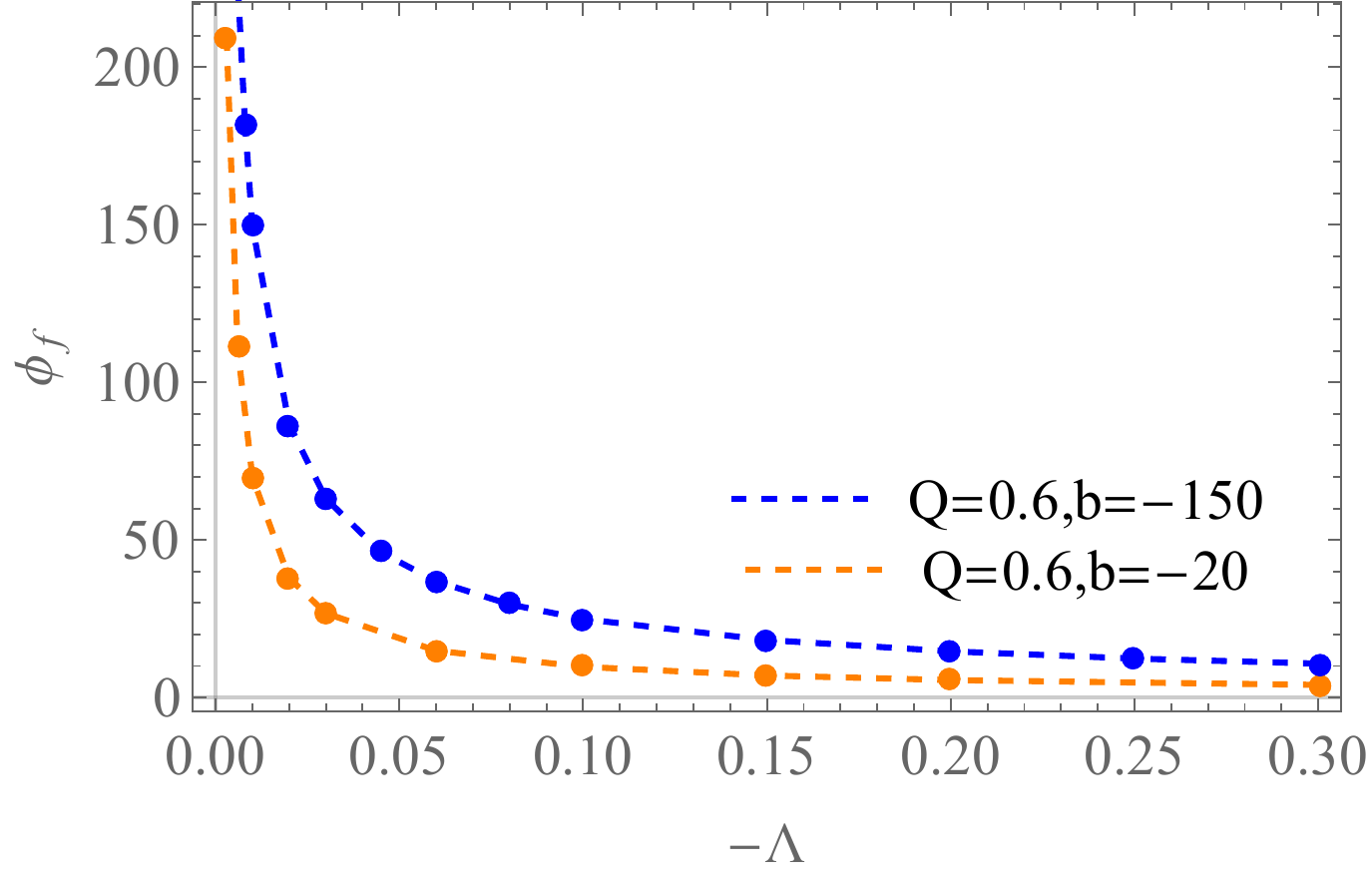}} &
{\footnotesize{}\includegraphics[width=0.32\textwidth]{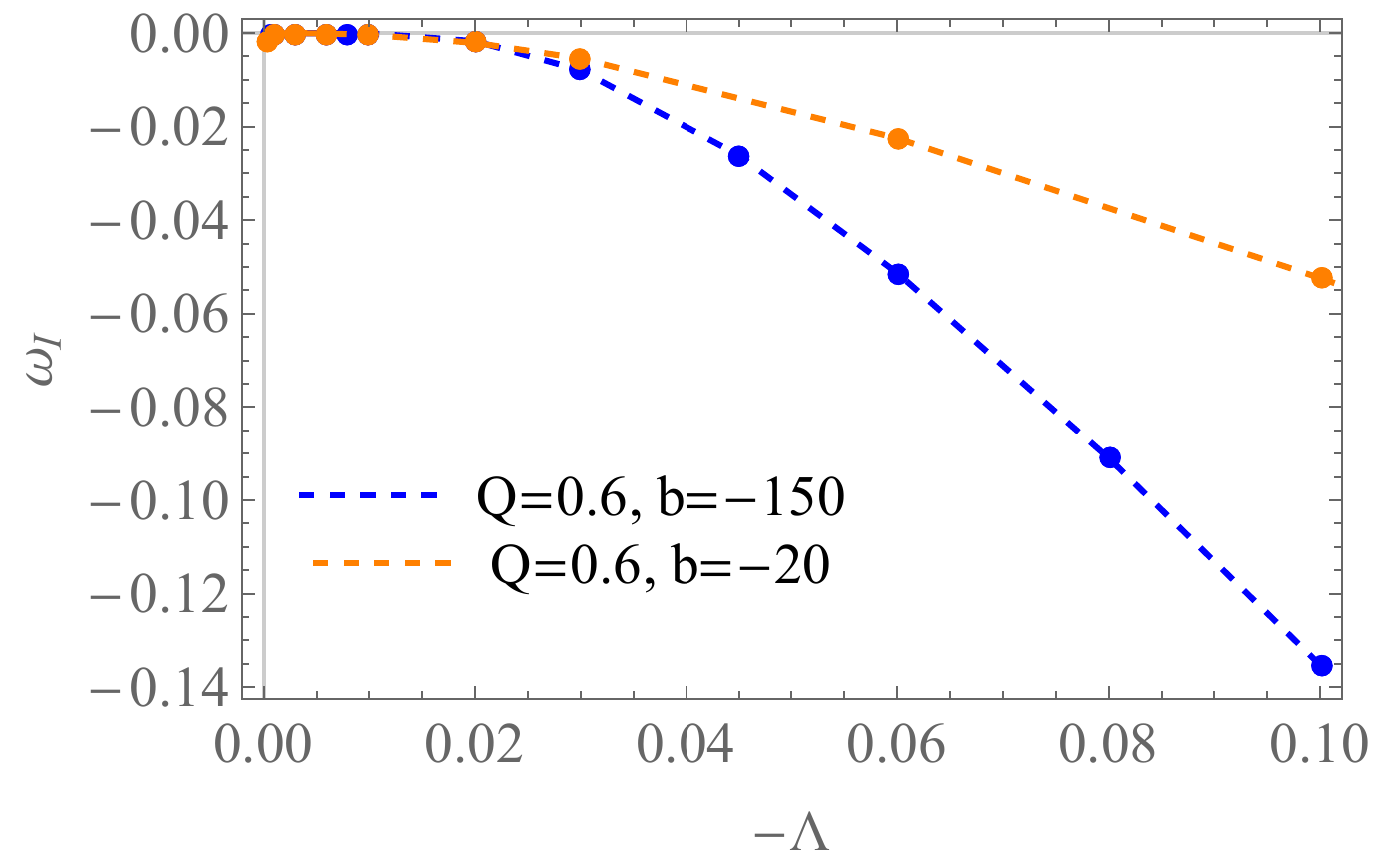}} & {\footnotesize{}\includegraphics[width=0.31\textwidth]{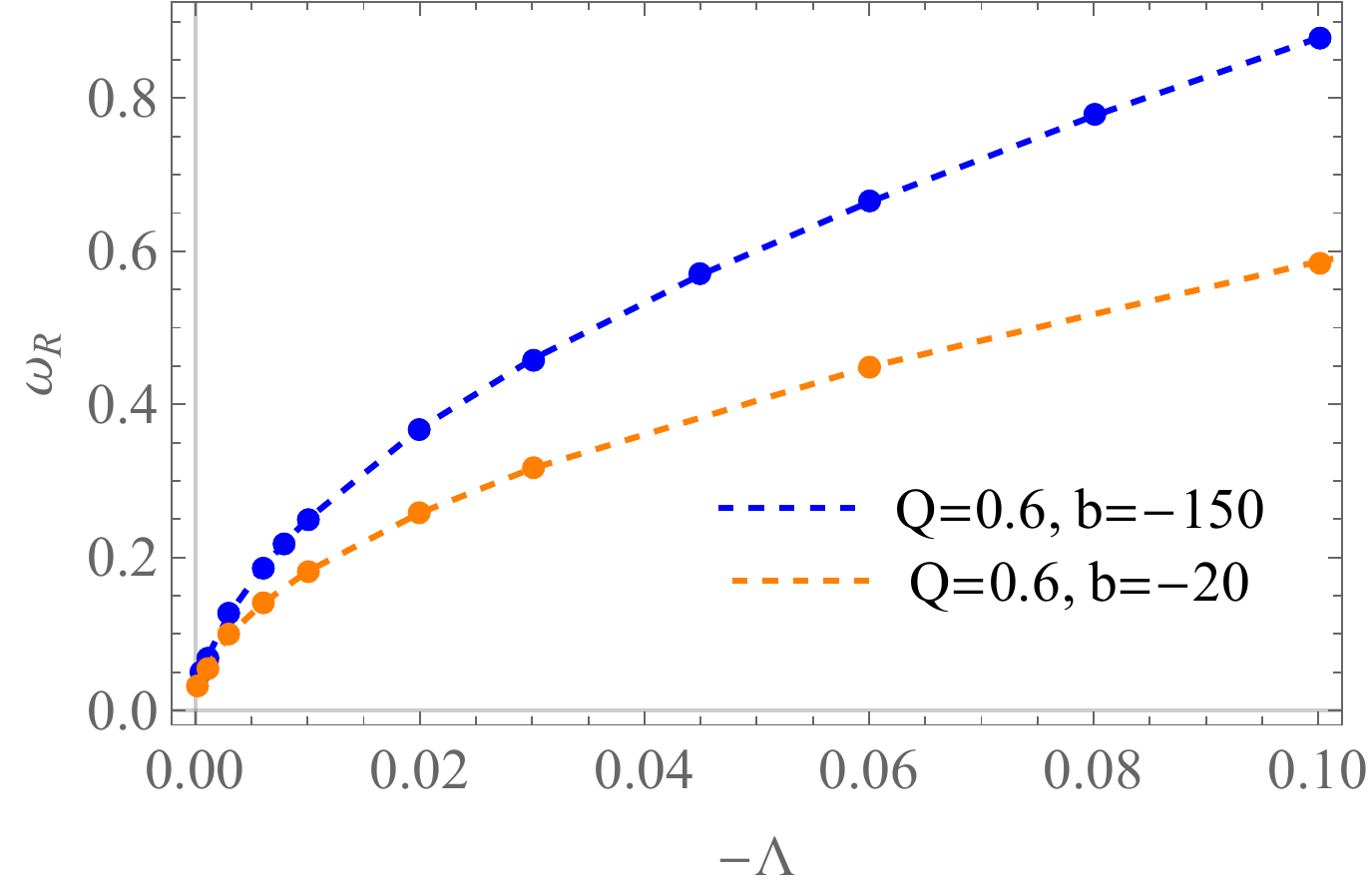}}\tabularnewline
\end{tabular}{\footnotesize\par}
\par\end{centering}
{\footnotesize{}\caption{\label{fig:LambdaPhi3}{\small{} }The evolution of $\phi_{3}$ (upper
left), the inverse final values of $\phi_{f}^{-1}$ of $\phi_{3}$
(upper right) and the complex frequencies of the dominant damping
modes of $\phi_{3}$ (lower panels) for various $\Lambda$.}
}{\footnotesize\par}
\end{figure*}

In this subsection, we fix $Q=0.6$ and choose $b=-20$ and $-150$ to study the effects of $\Lambda$ on the black hole evolution.
The AdS space looks like a potential well. The larger the $-\Lambda$, the deeper and narrower the potential well. Less energy of the Maxwell field is transferred to the scalar field. So  the final value $\phi_{f}$ decreases with $-\Lambda$  in the left panel of Fig.\ref{fig:LambdaPhi3}.
On the other hand, we find that the final value $\phi_{f}\propto\Lambda^{-1}$ as $\Lambda\to0$. This implies that the asymptotic solution in AdS spacetime can not be generalized straightforwardly to the asymptotic flat spacetime. In fact, the asymptotic expansion of the scalar field behaves as $\phi\sim c+\phi_{1}/r+\mathcal{O}(r^{-2})$ in asymptotic flat spacetime. The boundary condition should be changed to do the numerical calculations in asymptotic flat spacetime.

The complex frequencies of the dominant damping modes of $\phi_{3}$ are shown in the middle and right panels of Fig. \ref{fig:LambdaPhi3}. Both the imaginary part of the frequency $\omega_{I}$ and the real part $\omega_{R}$ tends to zero as $\Lambda\to0$. In AdS space, the smaller the cosmological constant, the flatter the potential well and the smaller the oscillating frequency of the scalar. The scalar needs more time to traverse the space and the system needs much more time to settle down when $-\Lambda$ is very small.
As $-\Lambda$ increases, $\omega_{I}$ decreases and $\omega_{R}$ increases. The system oscillates faster and damps faster. Note that unlike the case for EMS model, there is always a hairy black hole solution for large $-\Lambda$.

{\footnotesize{}}
{\footnotesize\par}

\begin{figure*}
\begin{centering}
{\footnotesize{}}%
\begin{tabular}{cc}
{\footnotesize{}\includegraphics[width=0.45\textwidth]{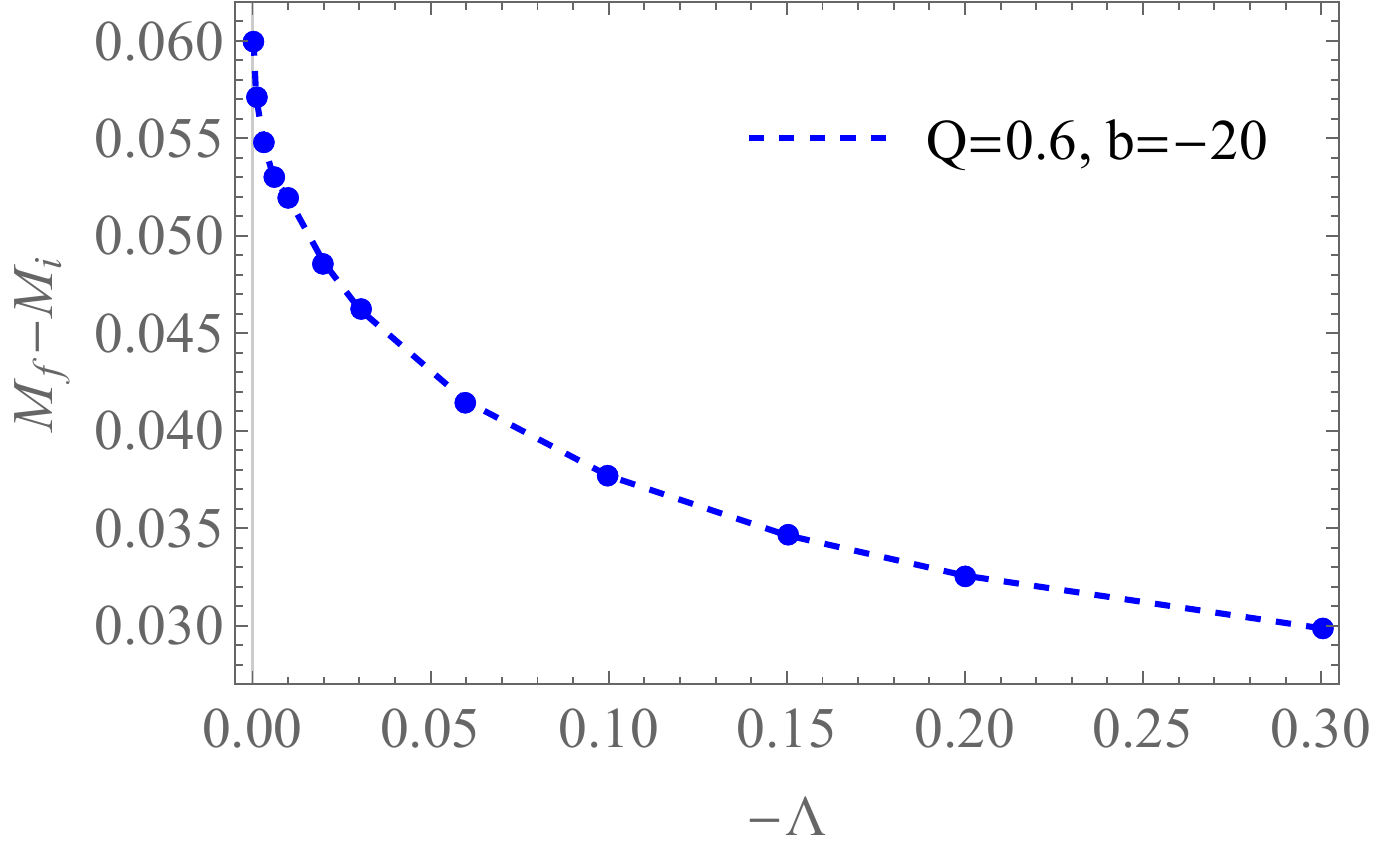}} & {\footnotesize{}\includegraphics[width=0.43\textwidth]{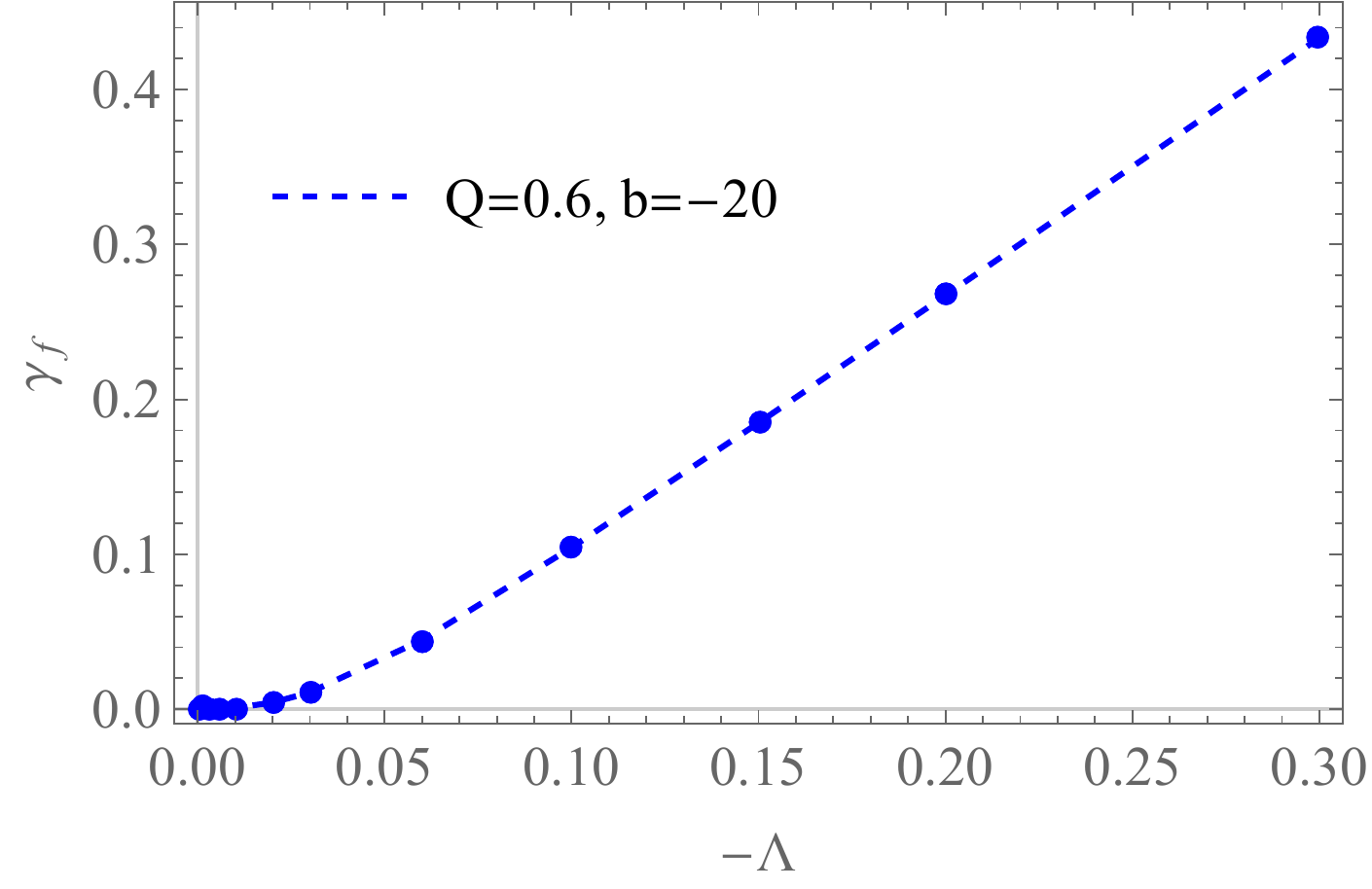}}\tabularnewline
\end{tabular}{\footnotesize\par}
\par\end{centering}
{\footnotesize{}\caption{\label{fig:LambdaMirr}{\small{} }The evolution of irreducible mass
$M_{0}$ (upper left), the increment of the irreducible mass $M_{f}-M_{i}$
(upper right), the evolution of $\ln(M_{f}-M_{0})$ (lower left) and
the growth rates $\gamma_{f}$ for various $\Lambda$ when $Q=0.6,b=-20$.}
}{\footnotesize\par}
\end{figure*}

The increment of the irreducible mass of the black hole is shown in the left panel of Fig.\ref{fig:LambdaMirr}. At initial times, $M_{0}$ increases abruptly and then increases as (\ref{eq:MirrExp}) at late times. The increment of the irreducible mass is larger for smaller $-\Lambda$, due to the relatively flat potential well which can accumulate more energy of the scalar that draws from the Maxwell field. The saturating rate $\gamma_{f}$ of the irreducible mass increases as $-\Lambda$ increases. This means that the system settles down faster as $-\Lambda$ increases.

{\footnotesize{}}
{\footnotesize\par}

\section{Summary and discussion}\label{sec:sum}

We studied the full nonlinear evolution of the spherical symmetric black holes under a small neutral scalar field perturbation in EMD theory. The equation of motion of the gauge field has no source and can be worked out directly. The free parameters are the black hole charge $Q$, the cosmological constant $\Lambda$, the coupling parameter $-b$ and the ADM mass of the system. The scalar hair can be represented by the coefficient $\phi_3$ of order $O(r^{-3})$ in the expansion near the infinity, which is determined by the evolution. We fixed the ADM mass $M=1$ to implement the dimensionless of the physical quantities.

We first studied the distribution of Misner-Sharp mass at late equilibrium time. It is constant to a relatively large radial distance, indicating that the scalar hair lies far away from the black hole for large coupling parameters. Then we show the final value of $\phi_3$. It increases monotonically with both the black hole charge and the coupling parameter. Unlike the case in the EMS model where the static hairy black hole solution exists only when $-b$ and $Q$ are large enough, there is always a static hairy black hole solution here. The evolution of $\phi_3$ resembles the quasinormal mode. The early time behavior of $\phi_3$ is closely related to the initial perturbation. Then it oscillates with damping amplitude and converges to the final value by a power-law. The decaying rate of the dominant decaying mode increases at first and then decreases with $-b$. The irreducible mass of the black hole increases abruptly at initial times and then saturates to the final value exponentially. {The saturating rate at late times is twice of the decaying rate of the dominant mode of $\phi_3$.} The system needs more time to settle down when $-b$ is intermediate. Note that we also simulated the evolution of singularity spacetime with a large charge to mass ratio. A horizon forms soon and hides the singularity, leaving a regular spacetime geometry outside the horizon.

We also studied the effects of the cosmological constant $\Lambda$ on evolution. As $\Lambda\to 0$, the final value of $\phi_{3}$ is proportional to $\Lambda^{-1}$. The boundary condition should be changed when implementing a numerical simulation in asymptotic flat spacetime. As $\Lambda\to 0$, both the complex frequency of the dominant mode and the saturating rate of the irreducible mass tends to zero. The system needs much more time to settle down for a small cosmological constant.

A natural generalization of this work is to study the evolution of black holes in asymptotic flat and dS spacetime. The boundary condition is different that there is not an effective potential barrier at infinity that could bounce the matter back to the bulk \cite{Hod:1996ar,Zhang:2015dwu}. Another interesting problem is to study the gravitational collapse or evolution of the black hole with a complex scalar field. The dynamics would be more rich and interesting \cite{Bosch:2016vcp,Chesler:2013lia,Dias:2016pma,Sanchis-Gual:2015lje,Chesler:2018txn}.

\section*{Acknowledgments}

Peng Liu would like to thank Yun-Ha Zha for her kind encouragement during this work. The authors thank De-chang Dai for his helpful discussion. This research is supported by National Key R\&D Program of China under Grant No.2020YFC2201400, and the Natural Science Foundation of China under Grant Nos.11690021, 11947067, 12005077, 11847055, 11905083, 11805083.

\end{document}